\documentclass[11pt]{article}

\usepackage{natbib}

\usepackage{graphicx}
\usepackage{here}
\usepackage[labelfont=bf,hang]{caption}
\usepackage[labelfont=sf]{subfig}
\usepackage{color}
\usepackage{bm}
\usepackage{amsmath, amsthm, amsfonts}

\usepackage{cases}
\usepackage{newtxtext}
\usepackage{fancyhdr}
\usepackage{longtable}
\usepackage{here}
\usepackage{booktabs}
\usepackage{algorithm}
\usepackage{algpseudocode}
\usepackage{multirow}
\usepackage{siunitx}
\usepackage{tikz}
\usetikzlibrary{arrows.meta, positioning}

\usepackage{enumitem}
	\setlist[itemize]{noitemsep}
	\setlist[description]{noitemsep}
	\setlist[enumerate]{noitemsep, font=\bf}

\usepackage{endnotes}

\usepackage[top=30truemm,bottom=30truemm,left=30truemm,right=30truemm]{geometry}

\usepackage[
	bookmarks=false,
	colorlinks=true,
	citecolor=blue,
	urlcolor=blue,
	linkcolor=blue,
 	pdfstartview=FitH,
]{hyperref}

\makeatletter
\renewcommand{\theHALG@line}{\thealgorithm.\arabic{ALG@line}}
\makeatother

\usepackage[capitalise,nameinlink]{cleveref}
\crefname{theo}{Theorem}{Theorems}
\crefname{prop}{Proposition}{Proposition}
\crefname{eq}{Eq.}{Eqs.}
\crefname{fig}{Fig.}{Figs.}
\crefname{tab}{Tab.}{Tabs.}
\crefname{sec}{Section}{Sections}

\def\Dt{\Delta t}

\let\OrigLparen\(
\let\OrigRparen\)
\def\({\left(}
\def\){\right)}
\def\[{\left[}
\def\]{\right]}
\def\<{\left\langle}
\def\>{\right\rangle}
\def\<<{\left\{}
\def\>>{\right\}}

\def\del{\partial}

\def\xx{\bm{x}}

\def\tthh{\bm{\theta}}

\def\cT{{\cal T}}

\def\->{$\rightarrow$}

\newcommand{\R}{\mathbb{R}}

\allowdisplaybreaks

\reversemarginpar

\begin{document}


\title{End-to-end differentiable network traffic simulation with dynamic route choice}
\author{
	Toru Seo\footnote{Institute of Science Tokyo; seo.t.aa@m.titech.ac.jp; The corresponding author}
}

\date{\today}
\maketitle

\begin{abstract}
	Optimization using network traffic models requires computing gradients of objective functions with respect to model parameters.
	However, derivation of such gradients has often been considered difficult or impractical due to their complexity and size.
	Conventional approaches rely on numerical differentiation or derivative-free methods that do not scale well with the parameter dimension, or on adjoint methods that require manual derivation for each specific model.
	This study proposes a novel end-to-end differentiable network traffic flow simulator based on automatic differentiation (AD), employing the Link Transmission Model (LTM) and a Dynamic User Optimum (DUO) route choice model.
	The LTM operates on continuous aggregate state variables through piecewise-linear min/max operations, which admit subgradients almost everywhere and thus require no smooth relaxation for AD.
	The DUO is also suitable for AD: although the shortest path search is itself discrete, the resulting diverge ratios at each node are continuous functions of per-destination vehicle counts and are thus differentiable.
	In order to demonstrate the capability of the proposed model, we solved a dynamic congestion toll optimization problem on the Chicago-Sketch dataset with approximately 2500 links, 1 million vehicles, a 3-hour duration, and \num{15000} decision variables.
	The proposed model successfully derived a high-quality solution in \num{3000} iterations, taking about 40 minutes.
	The simulator, implemented in Python and JAX, is released as open-source software named {\it UNsim} (\url{https://github.com/toruseo/UNsim}).
\end{abstract}

\begin{center}
\end{center}




\section{Introduction}

Network traffic flow models and simulation tools are fundamental to a wide range of traffic engineering and science tasks, from demand estimation and parameter calibration to network design optimization and real-time traffic control.
In general, these tasks share a common computational need: evaluating how simulation outputs (e.g., total travel time, link volumes, queue lengths) respond to changes in the simulation inputs (e.g., origin-destination demand, fundamental diagram parameters, toll, traffic signal configuration).
Efficient computation of this input--output sensitivity is therefore a central methodological challenge, particularly as the network scale and the dimension of the decision space grow.

Conventional approaches treat the simulator as a black box and employ derivative-free methods such as SPSA \citep{Spall1998spsa} and genetic algorithms \citep{goldberg1989ga}.
These methods are not very computationally efficient when the parameter dimension is large.
Surrogate-model or metamodel approaches can accelerate the search by approximating the derivatives using data-driven or analytical methods \citep{Osorio2015metamodel, song2017metamodel, Fu2024surrogate, dantsuji2024metamodelb}, but may introduce approximation errors or physical inconsistencies, especially when a data-driven approximation is employed.
An alternative is the adjoint method, which computes exact gradients analytically \citep{Reilly2015adjoint}.
However, the adjoint equations must be derived manually for each specific model, making the approach labor-intensive and difficult to maintain as the model evolves.

{\it Differentiable simulation} offers a way to combine the exactness of adjoint gradients with the generality of {\it Automatic Differentiation (AD)}.
In this approach, the simulator is written as a composition of differentiable operations, and the AD framework derives the gradient automatically (see Appendix \ref{apx_ad} for a brief overview of AD).
This paradigm has been applied successfully in other domains \citep{Hu2020difftaichi, Chen2018neuralode}.
In the traffic flow domain, \citet{Poole2016ad} applied AD to the METANET second-order macroscopic model \citep{papageorgiou2010metanet} for parameter calibration, and \citet{Son2022difftraf} proposed a differentiable hybrid simulator combining macroscopic and microscopic models.
However, these works have been limited to either single-link or simplified network settings, without endogenous route choice, which is essential for network-scale transportation analysis.

Extending differentiable simulation to the network scale with route choice poses a significant challenge: route decisions are inherently discrete (a vehicle either takes path A or path B), and discrete operations break the differentiability of the computation graph (see Appendix \ref{apx_ad} for a brief overview of computation graphs).
\citet{Andelfinger2022diffsim} and \citet{Makinoshima2026trafficsim} addressed this for agent-based simulators by replacing discrete branching with smooth approximations, but these relaxations alter the model behavior and may affect gradient quality, and they do not explicitly consider endogenous route choice either.

This study proposes a simple yet effective approach for traffic simulation with AD.
We observe that macroscopic traffic flow models based on the kinematic wave theory \citep{Lighthill1955LWR,Richards1956LWR,Newell1993kinematic1} operate on continuous aggregate state variables (cumulative vehicle counts), and that the piecewise-linear operations arising from demand--supply constraints and node models admit subgradients almost everywhere.\footnotemark{}
Furthermore, thanks to the elegant nature of cumulative vehicle counts, they yield not only macroscopic states but also {\it microscopic} vehicle trajectories \citep{makigami1971three}.
Based on this observation, we formulate a differentiable network traffic flow simulator based on the {\it Link Transmission Model (LTM)} \citep{Kuwahara2001duo, Yperman2006ltm,yperman2007ltm}, incorporating a {\it Dynamic User Optimum (DUO)} route choice model \citep{Ran1993duo,Kuwahara2001duo}, without introducing significant smoothing relaxation that compromises the models' properties.
The resulting simulator is {\it end-to-end differentiable} almost everywhere, meaning that gradients of an arbitrary simulation output (e.g., total travel time) with respect to each input (e.g., demand, capacity, toll) can be exactly and simultaneously computed via AD in a single operation.
We verified that the values obtained by AD are qualitatively and quantitatively reasonable using small toy networks.
Furthermore, we numerically confirmed the efficiency of the proposed method by applying it to a dynamic congestion pricing optimization problem based on a large-scale, realistic network called the Chicago-Sketch dataset.
The developed simulator, named {\it UNsim}, is implemented in Python \citep{python2022python} with JAX \citep{jax2018github} and released as open-source software.

\footnotetext{%
	Note that ``almost everywhere'' is a mathematical technical term with a rigorous definition.
	For example, ``$f(x)$ is differentiable almost everywhere'' means that, if the value of $x$ is chosen completely randomly, $f$ is differentiable with probability 1.
	However, there may exist non-differentiable points in $f$ with zero measure.
	A typical example is the ReLU function $\max \{0,x\}$ \citep{nair2010relu}, which is differentiable almost everywhere, but non-differentiable only at $x=0$.
	In practice, such measure-zero singularities cause no difficulty in gradient-based optimization, and ReLU is extensively used in machine learning.
}


The remainder of this paper is organized as follows.
\cref{sec_review} reviews the related literature.
\cref{sec_method} describes the simulation model and the differentiable formulation.
\cref{sec_numerical} presents numerical experiments that demonstrate the capability of the proposed model.
\cref{sec_conclusion} concludes the paper.

\section{Literature review}\label{sec_review}

\subsection{Macroscopic traffic flow models and network extensions}

The kinematic wave theory of \citet{Lighthill1955LWR} and \citet{Richards1956LWR} established the first-order macroscopic traffic flow model (LWR model).
\citet{Newell1993kinematic1,Newell1993kinematic2,Newell1993kinematic3} reformulated the LWR model under a triangular fundamental diagram (FD) using cumulative vehicle count curves $N(t,x)$, avoiding explicit shock tracking.
Based on this formulation, \citet{Kuwahara2001duo} developed a network-level dynamic traffic assignment (DTA) model with demand--supply functions, and \citet{Yperman2006ltm,yperman2007ltm} systematically proposed the discrete-time LTM.\footnotemark{}
The LTM provides an exact numerical solution to the LWR model and is computationally more efficient than the Cell Transmission Model (CTM) \citep{Daganzo1994ctm} owing to its coarser spatial discretization.

\footnotetext{
	In fact, these two models (i.e., \citet{Kuwahara2001duo} and \citet{Yperman2006ltm}) are almost equivalent.
	For the details, see \citet{Wada2017ltm}.
}

Node models govern flow transfer at junctions.
\citet{Daganzo1995ctm} proposed node models consistent with CTM.
\citet{Lebacque1996godunov} introduced the demand--supply framework for boundary conditions.
\citet{tampere2011node} established seven requirements that any first-order node model must satisfy and defined the generic class of node models.
\citet{flotterod2011node} proposed the Incremental Node Model (INM) for general multi-leg intersections, and \citet{Smits2015node} unified several node model families.

\subsection{Dynamic route choice}

Static traffic assignment is based on Wardrop's user equilibrium principle \citep{Wardrop1952equilibrium}, under which no traveler can reduce their travel time by changing their route.
Extending this principle to the time-varying, dynamic setting leads to the Dynamic User Equilibrium (DUE) \citep{Szeto2006dta}.
\citet{Friesz1993vi} formulated DUE as a variational inequality, where travelers experience equal and minimal actual travel times on all used routes for a given departure time.
Meanwhile, since DUE is computationally very challenging, its tractable proxy termed Dynamic User Optimum (DUO) is also widely utilized \citep{iryo2013dta}.
For example, \citet{Ran1993duo} proposed an instantaneous DUO model in which travelers choose routes based on the instantaneous travel times at the moment of departure.

Combining DUO with macroscopic network loading models yields a physically consistent DTA.
For example, \citet{Kuwahara2001duo} integrated DUO route choice with the LWR model to capture queue formation and spillback in many-to-many OD networks.
In the context of differentiable simulation, incorporating an endogenous route choice model is more challenging than using fixed turning rates, because the shortest path computation involves discrete decisions that depend on the simulation states.

\subsection{Calibration and optimization of traffic simulation models}

Calibration and optimization of DTA models has traditionally relied on derivative-free methods.
\citet{Spall1992spsa,Spall1998spsa} proposed the SPSA algorithm, which estimates the gradient using only two function evaluations regardless of the parameter dimension.
SPSA and its variants have been widely applied to DTA calibration \citep{Balakrishna2007calib, Lu2015wspsa}.
While these methods are effective for moderate-dimensional problems, their convergence slows as the number of parameters grows, motivating the search for gradient-based alternatives.
Other approaches use metamodel or surrogate model-based simulation optimization \citep[e.g.,][and references therein]{Osorio2015metamodel, song2017metamodel, Fu2024surrogate, dantsuji2024metamodelb}.
These methods typically treat the simulator as a black box, employ differentiable approximations such as neural networks, or use analytical differentiation of specific models.

Adjoint methods are also utilized for traffic flow optimization.
The adjoint method provides an analytical route to computing gradients of objective functions with respect to control parameters in PDE-constrained optimization.
\citet{Reilly2015adjoint} derived the discrete adjoint for networks of scalar conservation laws (LWR model) discretized by the Godunov scheme and applied it to coordinated ramp metering.
\citet{Kolb2026adjoint} extended the approach to multiclass traffic flow models with merges, diverges, and spillback.
These methods require manual derivation of the adjoint equations for each specific model formulation, which is labor-intensive and error-prone when the model is modified.

\subsection{Differentiable simulation and automatic differentiation}

Differentiable simulation has emerged as a paradigm in which the entire simulation is formulated as a composition of differentiable operations, enabling AD frameworks to compute gradients without manual derivation.
\citet{Griewank2008ad} is the standard reference for the theory and implementation of algorithmic differentiation.
\citet{Baydin2018adsurvey} provides a comprehensive survey of AD in machine learning.

Let us explain the meaning of end-to-end differentiable simulation by AD in the context of traffic simulation in more detail.
In principle, a traffic simulation can be expressed as the following equation:
\begin{align}
	y = J(\xx),
\end{align}
where the vector $\xx$ is the simulation input (e.g., link attributes, vehicle attributes, parameters of imposed traffic control scheme), the scalar $y$ is one of the outputs (e.g., total travel time, link traffic volume), and $J$ is the simulation method represented as a function.
If the function $J$ has {\it certain properties}, AD derives the exact value of the partial derivative of $J$ at specific $\xx$ (say $\xx_0$) with respect to all elements of $\xx$ simultaneously (i.e., $\nabla_{\xx} J |_{\xx=\xx_0}$) by using so-called {\it reverse-mode AD}.
This is the meaning of end-to-end differentiability in this context.
The computation time of reverse-mode AD is only a few times larger than that of the simulation itself, regardless of the dimension of $\xx$.
Therefore, AD is useful for wide-ranging applications such as gradient-based optimization and sensitivity analysis.

The challenge is that the traffic simulation function $J$ must satisfy several conditions to be end-to-end differentiable by AD.
First, the function $J$ should be continuous and differentiable almost everywhere.
This is a significant constraint for traffic simulations, as they often contain discrete agents and discrete route choice.
This can be relaxed by applying certain approximations \citep{Andelfinger2022diffsim,Makinoshima2026trafficsim}; however, such continuous approximations of inherently discrete concepts might introduce unpredictable errors.
Second, the function $J$ should be computable by a fixed number of mathematical operations.
In particular, operations such as for-loops and if-then rules, which frequently appear in standard traffic simulation methods, are not compatible with AD, because they may change the number of mathematical operations endogenously.
For further details on AD, see Appendix \ref{apx_ad}.

In the traffic domain, \citet{Poole2016ad} incorporated an AD method into the METANET second-order macroscopic model and calibrated FD parameters using resilient backpropagation.
This appears to be the earliest application of AD to a traffic flow model.
Meanwhile, \citet{Osorio2011network} reformulated a DTA model based on the LTM into a stochastic and differentiable form by using smoothed functions.
Subsequent studies employed this and similar models for metamodel-based optimization \citep{Osorio2015metamodel, chong2018metamodel}.
Although these methods do not use AD and cannot compute the partial derivatives with respect to all inputs in a scalable way, the idea of differentiable simulation is relevant.
In the context of AD, \citet{Son2022difftraf} proposed the first differentiable traffic simulator supporting both macroscopic and microscopic models, with applications to signal control.
\citet{Son2025diffidm} developed a parallelized differentiable simulator based on the Intelligent Driver Model.
\citet{Andelfinger2022diffsim} and \citet{Makinoshima2026trafficsim} addressed the differentiability of agent-based simulations by replacing discrete branching with smooth approximations.
\citet{du2025modeling} proposed a differentiable simulation model for passenger flow in a metro network, in which a logit model was used to make passenger choice differentiable.
\citet{li2026network} proposed a differentiable static traffic assignment model based on a day-to-day model.

In summary, existing differentiable traffic simulators with AD have focused on either microscopic models or simplified macroscopic settings.
To the author's knowledge, no study integrates end-to-end differentiability with endogenous dynamic route choice at the network scale.
The most relevant and important prior works are \citet{Osorio2011network, Osorio2015metamodel, chong2018metamodel}, which derived analytical gradients of macroscopic and metamodel formulations and used them for gradient-based optimization of congested networks, although they did not use AD.
These contributions have established that differentiability is a powerful property for traffic optimization.

The present work builds on this perspective and complements it from a different angle: instead of deriving gradients analytically, we construct the entire dynamic network loading and route choice procedure as a composition of operations compatible with reverse-mode AD.
As a result, the gradient with respect to all decision variables is obtained in a single backward pass at a cost independent of the parameter dimension.
This is particularly suitable for high-dimensional optimization problems, such as dynamic congestion pricing on large-scale networks.

\section{Methodology}\label{sec_method}

\begin{figure}[htp]
	\centering
	\includegraphics[clip, width=0.99\hsize]{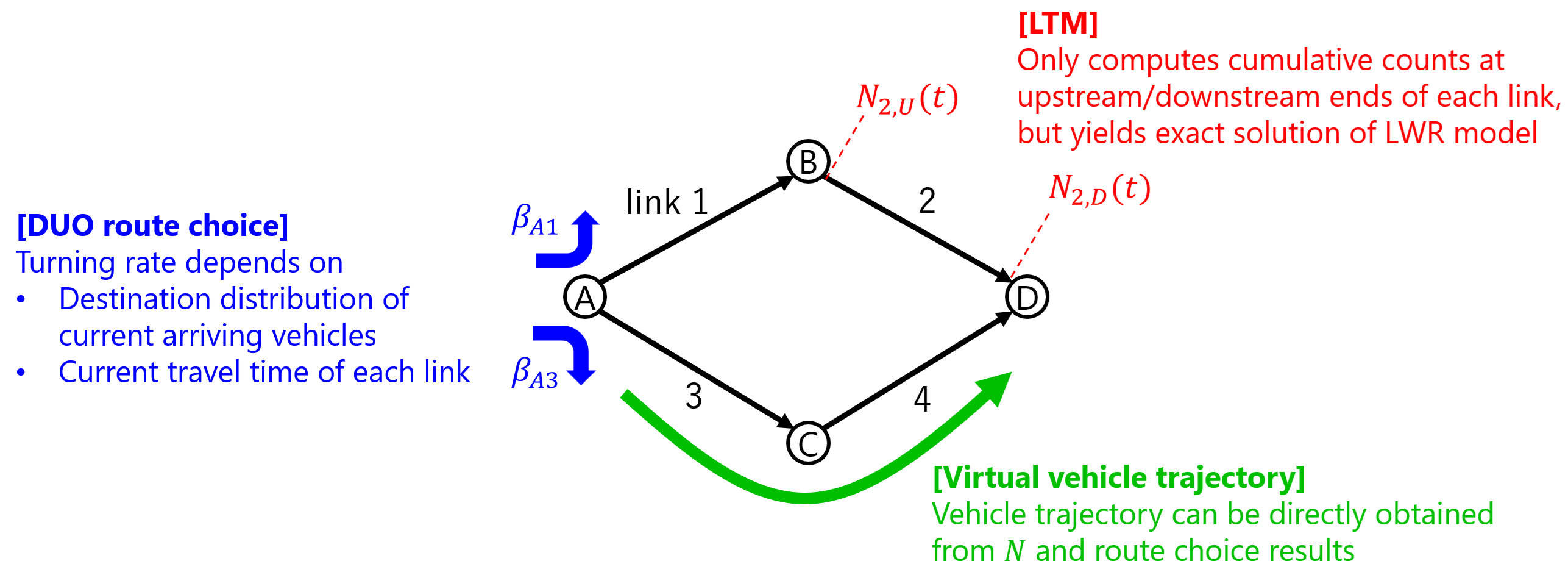}
	\caption{The simulation framework combining LTM traffic flow model and DUO route choice.}
	\label{LTM_DUO}
\end{figure}

In this section, we present the formulation of the proposed end-to-end differentiable macroscopic traffic flow simulator.
The simulator is based on the Link Transmission Model (LTM), combined with a Dynamic User Optimum (DUO)-type endogenous route choice model, and it is formulated such that every operation is differentiable with respect to the model parameters.
We first describe the parameters and state variables of the proposed model, and the meaning of a differentiable simulator in the traffic simulation context (\cref{sec_network}).
Then, we overview the base models (\cref{sec_base_model}), which are direct applications of the existing works.
Finally, we describe the differentiable formulation of these base models (\cref{sec_diff}), which is the key contribution of this work.
\Cref{LTM_DUO} illustrates the conceptual framework of the proposed model.

\subsection{Parameters and state variables, and what the differentiable simulator means}\label{sec_network}

A road network is represented as a directed graph $\mathcal{G} = (\mathcal{N}, \mathcal{L})$, where $\mathcal{N}$ denotes a set of nodes and $\mathcal{L}$ denotes a set of directed links.
Each link $l \in \mathcal{L}$ connects an upstream node to a downstream node.
For each node $\nu \in \mathcal{N}$, we define $\mathcal{L}_\nu^{\mathrm{in}}$ and $\mathcal{L}_\nu^{\mathrm{out}}$ as the sets of incoming and outgoing links, respectively.

The simulation is governed by a parameter vector $\tthh$ (i.e., input) and produces state variables (including output) that evolve over time.
The parameter vector $\tthh$ consists of:
\begin{itemize}
	\item For each link $l$: free-flow speed $u_l$, jam density $\kappa_l$, and backward wave speed $w_l$,
	\item For each node $\nu$: merge priorities $\alpha_l$ for each inlink $l$,
	\item OD demand: flow rate $q_{rs}(t)$ from origin $r$ to destination $s$ at time $t$.
\end{itemize}

The primary state variables are the cumulative vehicle counts at the upstream and downstream boundaries of link $l$ on time $t$: $N_{l,U}(t)$ and $N_{l,D}(t)$, respectively.
Under the DUO route choice model, per-destination cumulative counts $N_{l,U}^s(t)$ and $N_{l,D}^s(t)$, where $s$ denotes the destination, are additionally maintained.
The simulation proceeds in discrete timesteps of width $\Dt$.
The simulation duration is $T_{\max}$, and the total number of timesteps is $T_S = T_{\max}/\Dt$.
The time index $t$ refers to the $t$-th timestep (i.e., $t \cdot \Dt$ seconds).

In the proposed model, any state variable is differentiable with respect to an arbitrary parameter almost everywhere.
Furthermore, any differentiable function of the state variables is also differentiable.
For example, the traffic volume of a link over a certain duration can be defined as
\begin{align}
	n = N_{l,D}(t+\Delta t)-N_{l,D}(t).
\end{align}
Then, by using AD, we can directly and simultaneously compute values of $\del n/\del u_l$, $\del n/\del q_{rs}(t)$, and so on by just performing one forward simulation and one AD operation.
The computation time for an AD operation is several times (usually 2--5 times) longer than that of the forward simulation alone and does not depend on the number of parameters.
Similarly, we can also define the total travel time $TTT$ based on the state variables.
Partial derivatives such as $\del TTT/\del u_l$ and $\del TTT/\del q_{rs}(t)$ can also be computed.

\subsection{Base models}\label{sec_base_model}

In this section, we overview the employed base models for simulation, namely, LTM \citep{Kuwahara2001duo, Yperman2006ltm}, INM \citep{flotterod2011node}, and DUO \citep{Kuwahara2001duo}.
Since they have been proposed in existing studies, this section provides only the information necessary to understand these models, along with the specifications adopted in the proposed simulation model.
For the details on these models, refer to the respective original articles.

\subsubsection{Link model: Link Transmission Model}\label{sec_ltm}

The link model is based on the LTM \citep{Kuwahara2001duo, Yperman2006ltm,yperman2007ltm}, which provides an exact numerical solution to the LWR model \citep{Lighthill1955LWR, Richards1956LWR} under a triangular FD.

\begin{figure}[htp]
	\centering
	\includegraphics[clip, width=0.5\hsize]{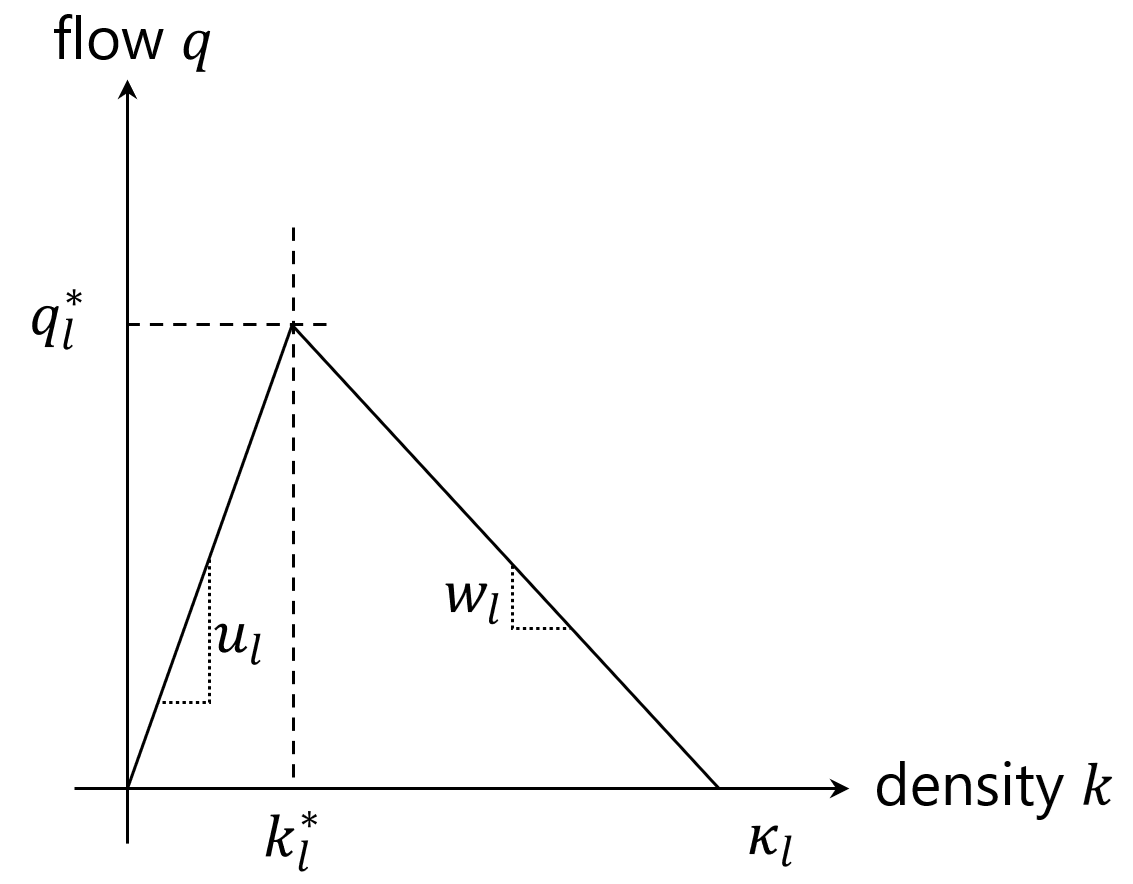}
	\caption{Triangular FD.}
	\label{fig_fd}
\end{figure}

Each link $l$ is characterized by a triangular FD (\cref{fig_fd}) with three parameters: free-flow speed $u_l$, traffic capacity $q_l^*$, and jam density $\kappa_l$.
The flow--density relation is
\begin{numcases}{Q_l(k) =}
	u_l k	&	if $k \leq k_l^*$,	\label{eq_fd_free}\\
	w_l(\kappa_l - k) & if $k > k_l^*$,	\label{eq_fd_cong}
\end{numcases}
where $k$ denotes the traffic density, $k_l^* = q_l^*/u_l$ is the critical density, and $w_l=q_l^*/(\kappa_l-k_l^*)$ is the backward wave speed.
The corresponding speed--density relation is
\begin{numcases}{V_l(k) = Q_l(k)/k =}
	u_l	&	if $k \leq k_l^*$,	\label{eq_vd_free}\\
	w_l(\kappa_l - k) / k & if $k > k_l^*$.	\label{eq_vd_cong}
\end{numcases}

The LTM uses the cumulative vehicle counts (i.e., the number of vehicles that passed the location from a certain reference time to the current time) at the upstream and downstream ends of each link as its state variables.
They are denoted as $N_{l,U}(t)$ for upstream and $N_{l,D}(t)$ for downstream, respectively, of link $l$ at time $t$.

The LTM is derived from Newell's simplified kinematic wave (KW) theory \citep{Newell1993kinematic1,Newell1993kinematic2,Newell1993kinematic3}, which determines the cumulative vehicle count at any time--space point $(t, x)$ within a link as
\begin{align}
	N_l(t, x) = \min\left\{
		N_{l,U}\!\(t - \frac{x}{u_l}\),~
		N_{l,D}\!\(t - \frac{d_l - x}{w_l}\) + \kappa_l(d_l - x)
	\right\},		\label{eq_newell}
\end{align}
where $d_l$ is the length of link $l$, and $x$ is the distance from the upstream boundary.
The first argument corresponds to the free-flow characteristic, and the second to the congestion characteristic.

At each timestep $t$, the LTM computes the {\it demand} $D_l(t)$ and {\it supply} $S_l(t)$ of each link.
The demand represents the maximum flow rate that can depart from the downstream boundary, while the supply represents the maximum flow rate that the link can accept at its upstream boundary.
By applying the Newell formula (\cref{eq_newell}) to each end ($x=0$ or $d_l$), they are computed as
\begin{align}
	D_l(t) &= \min\left\{ \frac{N_{l,U}(t + \Dt - d_l/u_l) - N_{l,D}(t)}{\Dt},~ q_l^* \right\},		\label{eq_demand}\\
	S_l(t) &= \min\left\{ \frac{N_{l,D}(t + \Dt - d_l/w_l) + \kappa_l d_l - N_{l,U}(t)}{\Dt},~ q_l^* \right\}.		\label{eq_supply}
\end{align}
The non-negativity constraints $D_l(t) \geq 0$ and $S_l(t) \geq 0$ are also enforced.

\begin{figure}[htp]
	\centering
	\includegraphics[clip, width=0.9\hsize]{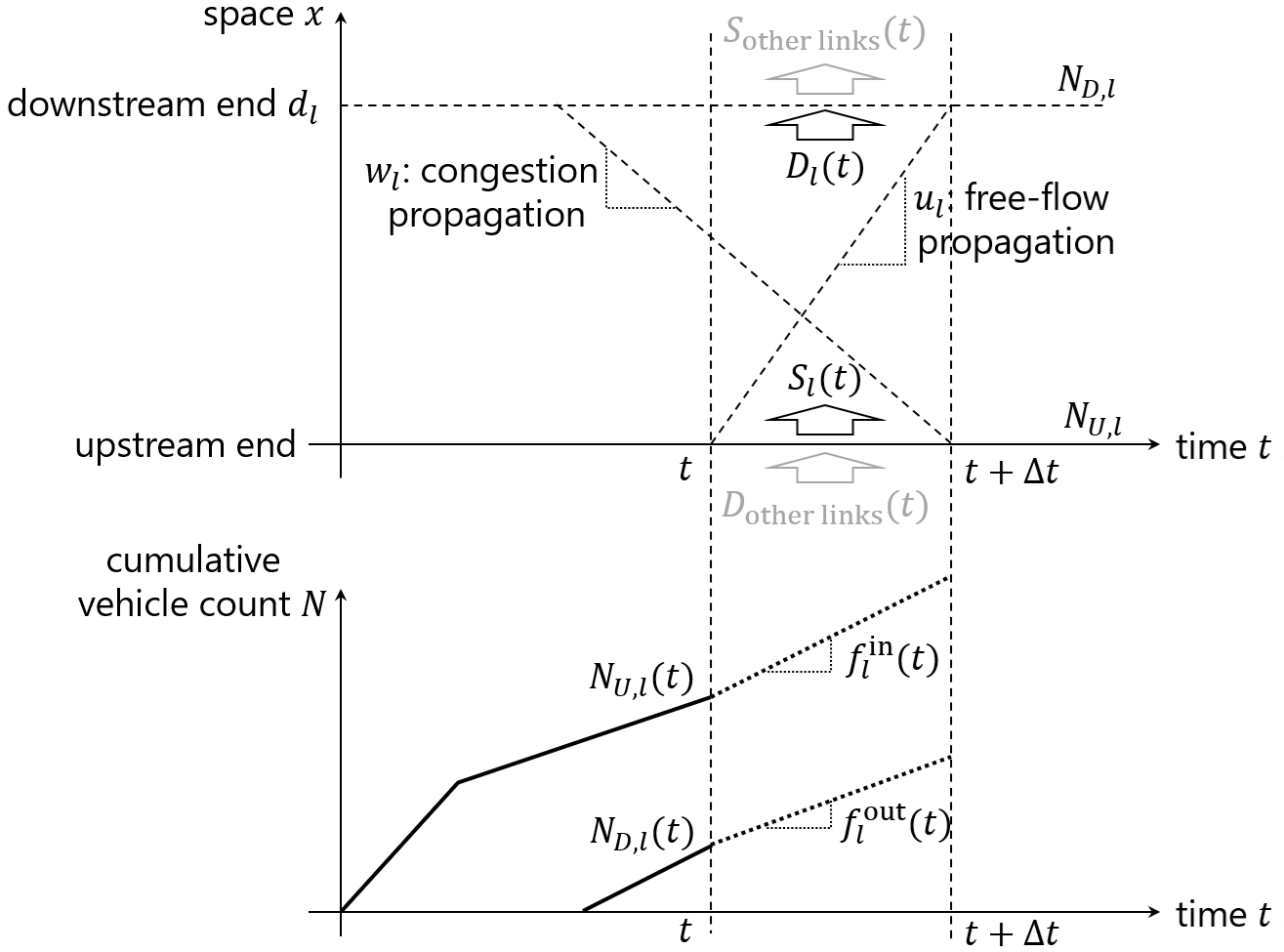}
	\caption{Mechanism of LTM. Top: time--space diagram on link $l$. Bottom: its cumulative curve plot. Link $l$ determines its demand $D_l$ and supply $S_l$ considering its traffic state. Then, the node model determines inflow $f_l^{\mathrm{in}}$ and outflow $f_l^{\mathrm{out}}$ considering demand and supply of connected links.}
	\label{schematic_ltm}
\end{figure}

The realized flow that enters link $l$ during time $[t, t+\Delta t)$ is denoted as $f_l^{\mathrm{in}}(t)$, and the flow that leaves the link is denoted as $f_l^{\mathrm{out}}(t)$.
These values are determined by the node model described in \cref{sec_node}, considering the interactions among connected links.
Then, the cumulative counts are updated as
\begin{align}
	N_{l,U}(t + \Dt) &= N_{l,U}(t) + \Dt \cdot f_l^{\mathrm{in}}(t),	\label{eq_update_U}\\
	N_{l,D}(t + \Dt) &= N_{l,D}(t) + \Dt \cdot f_l^{\mathrm{out}}(t).	\label{eq_update_D}
\end{align}
See \cref{schematic_ltm} for an illustration of the mechanism of the LTM.

The timestep width must satisfy $\Dt \leq d_l / u_l$ for all links to ensure numerical stability, which is the Courant--Friedrichs--Lewy (CFL) condition for the LTM \citep{yperman2007ltm}.
However, this width is generally significantly larger than that of other models such as CTM and microscopic car-following models.

\subsubsection{Virtual vehicle trajectories in LTM}\label{sec_virtual_vehicle}

\begin{figure}[htp]
	\centering
	\includegraphics[clip, width=0.8\hsize]{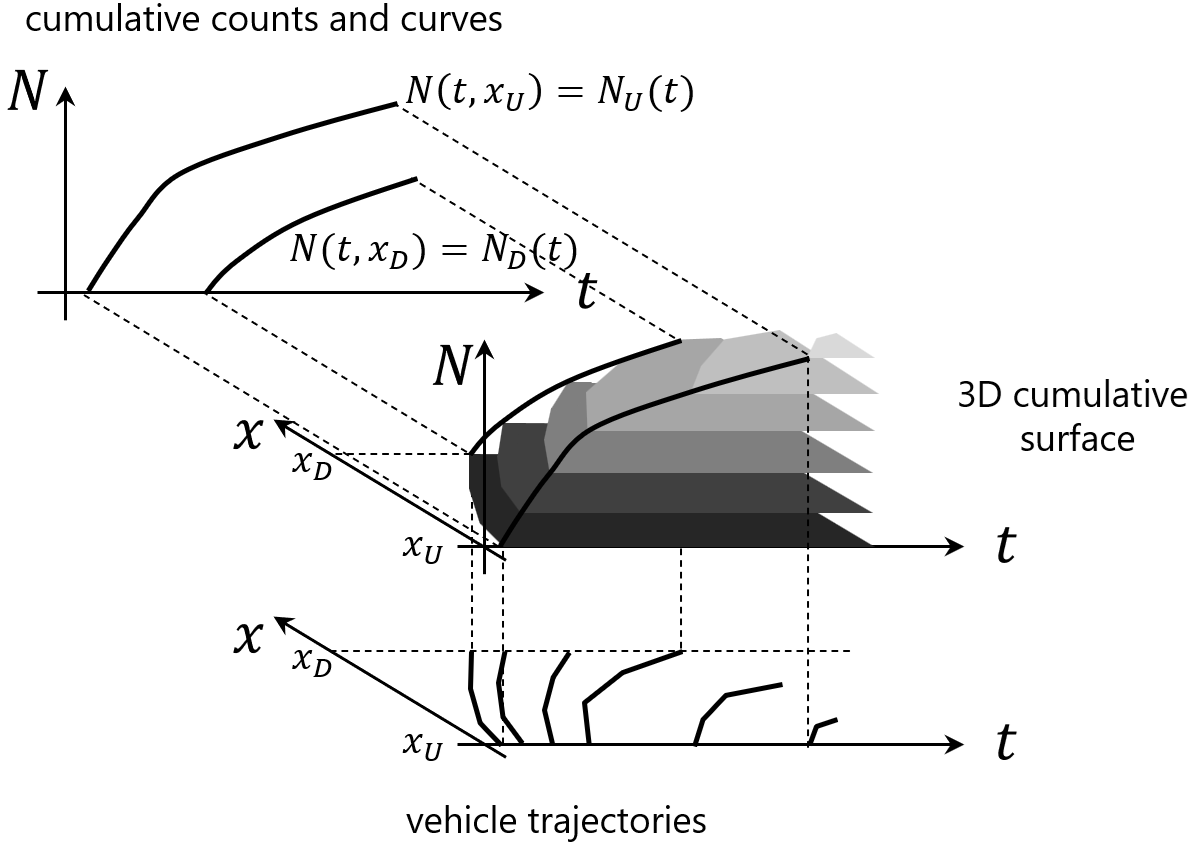}
	\caption{Vehicle trajectories and cumulative counts (adapted from \citet{seo2023book_en}).}
	\label{cumulative_count_3d}
\end{figure}

A notable property of the cumulative count representation is that it yields not only macroscopic traffic states but also individual vehicle trajectories \citep{makigami1971three}.
Since the cumulative count $N_{l,U}(t)$ records the total number of vehicles that have entered link $l$ by time $t$, a vehicle entering at time $t_{\mathrm{enter}}$ is assigned the index $N = N_{l,U}(t_{\mathrm{enter}})$.
Under the first-in first-out (FIFO) assumption, this vehicle is also the $N$-th vehicle to exit the link; therefore, its exit time is determined by the time at which the downstream cumulative count reaches the same value:
\begin{align}
	t_{\mathrm{exit}} = \max\left\{ t_{\mathrm{enter}} + \frac{d_l}{u_l},~ N_{l,D}^{-1}(N) \right\},	\label{eq_exit_time}
\end{align}
where $N_{l,D}^{-1}(N)$ denotes the time at which $N_{l,D}(t) = N$.
The first argument represents the free-flow constraint (the vehicle cannot traverse the link faster than $u_l$), and the second represents the queuing constraint (the vehicle must wait for preceding vehicles to depart).
\Cref{cumulative_count_3d} illustrates the relation between vehicle trajectories and cumulative counts.

Regarding network-level trajectories, for a path $\mathcal{P} = (l_1, l_2, \ldots, l_P)$, the travel time of a virtual vehicle departing at time $t_0$ is obtained by chaining \cref{eq_exit_time} over successive links:
\begin{align}
	t_p = \max\left\{ t_{p-1} + \frac{d_{l_p}}{u_{l_p}},~ N_{l_p,D}^{-1}\!\(N_{l_p,U}(t_{p-1})\) \right\}, \quad p = 1, \ldots, P,	\label{eq_chain_tt}
\end{align}
where $t_p$ denotes the exit time from the $p$-th link.
The total travel time is $t_P - t_0$.

The path $\mathcal{P}$ itself can be determined by a time-dependent shortest path algorithm, following the DUO route choice model explained later.
Since the LTM satisfies the FIFO property on every link (a vehicle entering earlier always exits earlier), Dijkstra's algorithm is applicable with the actual congestion-dependent link exit time (\cref{eq_exit_time}) as the edge cost.
Starting from the origin at time $t_0$, the algorithm expands nodes in order of earliest arrival time, using $t_{\mathrm{exit}}$ computed from the simulation's cumulative counts to evaluate each edge.
The resulting path is optimal in the sense that no other path yields an earlier arrival at the destination for the given departure time.

This formulation provides microscopic trip-level information (individual travel times and paths) from a macroscopic simulation, without computing individual vehicles explicitly.

\subsubsection{Node model: Incremental Node Model}\label{sec_node}

\begin{figure}[htp]
	\centering
	\includegraphics[clip, width=0.33\hsize]{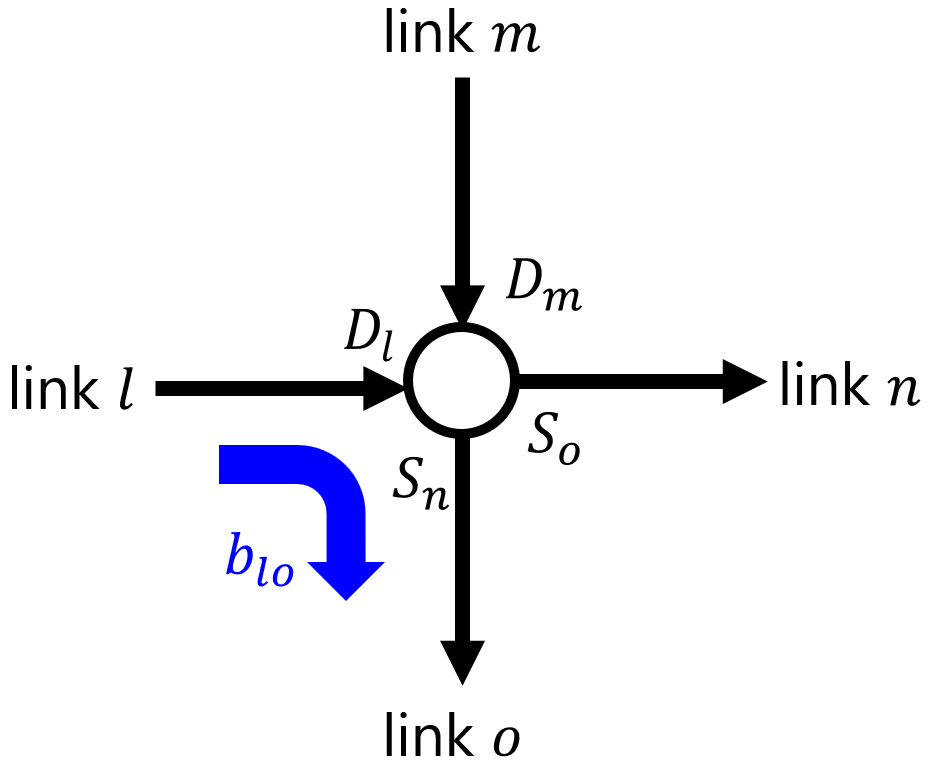}
	\caption{Node with multiple incoming and outgoing links.}
	\label{node_model}
\end{figure}

The node model determines the flow transfer at each node by considering the demand and supply of connected links (\cref{node_model}) to derive physically reasonable flow transfer and congestion spillback \citep{tampere2011node}.
We employ the INM \citep{flotterod2011node}.
The INM determines flows through an iterative procedure that incrementally allocates flow along a priority-weighted direction until a demand or supply constraint is reached.
In this section, the time variable $t$ is omitted for the sake of brevity (e.g., $D_l(t)$ is denoted as $D_l$).

Let $B_\nu = [b_{lo}]$ denote the turning fraction matrix, where $b_{lo}$ is the fraction of outflow from inlink $l$ directed to outlink $o$, and let $\alpha_l$ denote the merge priority of inlink $l$.
The value of $B_\nu$ is obtained from the route choice model.
The INM initializes inflow allocations $\hat{q}_l = 0$ for all $l$ and outflow allocations $\hat{q}_o = 0$ for all $o$, then iterates the following steps:
\begin{enumerate}
	\item Identify the set of {\it active inlinks} $\mathcal{D}^{\uparrow}$: inlinks $l$ where $\hat{q}_l < D_l$ and no outlink $o$ with $b_{lo} > 0$ is at capacity ($\hat{q}_o < S_o$).
	\item Identify the set of {\it active outlinks} $\mathcal{D}^{\downarrow}$: outlinks $o$ where $\hat{q}_o < S_o$ and at least one active inlink feeds them.
	\item If $\mathcal{D}^{\uparrow} = \emptyset$ (empty), terminate.
	\item Compute the priority-weighted direction: $\phi_l^{\mathrm{in}} = \alpha_l$ for $l \in \mathcal{D}^{\uparrow}$ (0 otherwise), and $\phi_o^{\mathrm{out}} = \sum_l b_{lo} \phi_l^{\mathrm{in}}$.
	\item Compute the maximum step size:
		\begin{align}
			\theta = \min\left\{
				\min_{l \in \mathcal{D}^{\uparrow}} \frac{D_l - \hat{q}_l}{\phi_l^{\mathrm{in}}},~
				\min_{o \in \mathcal{D}^{\downarrow}} \frac{S_o - \hat{q}_o}{\phi_o^{\mathrm{out}}}
			\right\}.		\label{eq_inm_step}
		\end{align}
	\item Update: $\hat{q}_l \leftarrow \hat{q}_l + \theta \phi_l^{\mathrm{in}}$ and $\hat{q}_o \leftarrow \hat{q}_o + \theta \phi_o^{\mathrm{out}}$.
\end{enumerate}
The procedure terminates in at most $|\mathcal{L}_\nu^{\mathrm{in}}| + |\mathcal{L}_\nu^{\mathrm{out}}| + 1$ iterations, as at least one constraint becomes binding at each step.
The final values $\hat{q}_l$ and $\hat{q}_o$ are the allocated outflow and inflow rates, respectively.

\subsubsection{Route choice model: Dynamic User Optimum}\label{sec_route}

The node model in the previous section requires diverge ratios $\beta_{\nu o}$ (or turning fractions $b_{lo}$) as exogenous inputs.
They are determined endogenously by a route choice model considering vehicles' minimum-travel-time path-search-like behavior.

We employ the DUO model as the route choice model.
Under the DUO assumption, vehicles at each node select the outgoing link that lies on the instantaneous shortest path to their destination, approximating equilibrium routing.
This requires tracking per-destination flows and computing shortest paths at some route choice update interval.
We provide two methods for computing the instantaneous link travel time, which differ in their treatment of within-link traffic heterogeneity.

The first method computes a single average density for the entire link from the boundary cumulative counts:
\begin{align}
	\bar{k}_l(t) = \frac{N_{l,U}(t) - N_{l,D}(t)}{d_l}.	\label{eq_tt_avg_k}
\end{align}
The corresponding speed is $\bar{v}_l(t) = V_l(\bar{k}_l(t))$ (\cref{eq_vd_free,eq_vd_cong}), and the travel time is
\begin{align}
	\tau_l(t) = \frac{d_l}{V_l(\bar{k}_l(t))}.	\label{eq_tt_avg}
\end{align}
This method is computationally inexpensive, as it requires only the boundary values that are already maintained by the LTM.
However, it assumes uniform density along the link and thus cannot distinguish, for example, a link that is half-congested from one with uniformly moderate density.

The second method captures within-link heterogeneity by dividing the link into $M$ equally spaced segments of width $\Delta x = d_l / M$.
For each segment $i = 1, \ldots, M$ with boundaries $x_{i-1} = (i{-}1)\Delta x$ and $x_i = i\,\Delta x$, the segment density is computed from the Newell $N$-curves (\cref{eq_newell}) at the boundaries:
\begin{align}
	k_l^{(i)}(t) = \frac{N_l(t, x_{i-1}) - N_l(t, x_i)}{\Delta x}.	\label{eq_tt_density}
\end{align}
The segment speed is $v_l^{(i)}(t) = V_l(k_l^{(i)}(t))$ (\cref{eq_vd_free,eq_vd_cong}), and the instantaneous travel time is the sum of the segment traversal times:
\begin{align}
	\tau_l(t) = \sum_{i=1}^{M} \frac{\Delta x}{V_l(k_l^{(i)}(t))}.	\label{eq_tt}
\end{align}
This method evaluates the Newell $N$-curve at interior points of the link, thereby resolving the spatial distribution of congestion (e.g., a queue occupying only the downstream portion of the link).
It is more computationally expensive than the average-density method, as it requires $2M$ evaluations of \cref{eq_newell} per link.

Given the link travel times $\{\tau_l(t)\}$, the shortest path tree from each destination $s$ is computed by a reverse Bellman--Ford algorithm \citep{ford1956shortestpath,bellman1958shortest}.
For each node $\nu$, let $\pi_\nu^s$ denote the next link on the shortest path toward destination $s$.
The shortest paths are recomputed periodically (every $\Delta t_{\mathrm{route}}$ timesteps, where $\Delta t_{\mathrm{route}}$ denotes the route update interval) rather than at every timestep, balancing accuracy and computational cost.

The diverge ratios at each node are then determined by aggregating the per-destination shortest path indicators, weighted by the current per-destination traffic volume.
Let $n_{l}^s(t) = N_{l,U}^s(t) - N_{l,D}^s(t)$ denote the number of destination-$s$ vehicles on link $l$ at time $t$.
For each node $\nu$ and outlink $o \in \mathcal{L}_\nu^{\mathrm{out}}$, the diverge ratio is
\begin{align}
	\beta_{\nu o}(t) = \frac{\sum_s \omega_\nu^s(t) \cdot \mathbf{1}[\pi_\nu^s = o]}
	                        {\sum_s \omega_\nu^s(t)},	\label{eq_duo_beta}
\end{align}
where $\omega_\nu^s(t) = \sum_{l \in \mathcal{L}_\nu^{\mathrm{in}}} n_l^s(t)$ for intermediate nodes, and $\omega_\nu^s(t) = q_{rs}(t)$ (the OD demand rate from origin $r = \nu$) for origin nodes.
In other words, the fraction of flow directed to outlink $o$ equals the fraction of vehicles (across all destinations) whose shortest path passes through $o$.

In order to evaluate \cref{eq_duo_beta}, the DUO model maintains per-destination cumulative counts $N_{l,U}^s(t)$ and $N_{l,D}^s(t)$ for each link and destination.
At the upstream boundary, the per-destination inflow is determined by the routing decision (\cref{eq_duo_beta}).
At the downstream boundary, the per-destination outflow follows the FIFO principle:
\begin{align}
	f_{l,s}^{\mathrm{out}}(t) = f_l^{\mathrm{out}}(t) \cdot \frac{N_{l,U}^s(t)}{N_{l,U}(t)},	\label{eq_fifo}
\end{align}
where $f_{l,s}^{\mathrm{out}}(t)$ denotes the outflow rate of destination-$s$ vehicles from link $l$.
This allocates the aggregate outflow proportionally to the destination composition of cumulative arrivals.

\subsection{End-to-end differentiable formulation}\label{sec_diff}

This section describes how the combined LTM and DUO simulation described above is formulated as an end-to-end differentiable model, enabling exact gradient computation via automatic differentiation (AD).
For code implementation, we use Python and JAX.

\subsubsection{Why the LTM is suitable for AD-based network traffic simulation}\label{sec_diff_why}

Agent-based (Lagrangian) traffic simulators track individual vehicles through inherently discrete events (e.g., route decisions) that break the differentiability of the simulation mapping.
Obtaining gradients from such models requires variance-prone estimators such as policy gradients or the straight-through estimator \citep{Makinoshima2026trafficsim}.

The LTM, in contrast, operates on continuous aggregate state variables, namely the cumulative vehicle counts $N_{l,U}(t)$ and $N_{l,D}(t)$, and all operations reduce to arithmetic and piecewise-linear $\min$/$\max$ functions.
Since $\min$ and $\max$ are differentiable almost everywhere (with standard subgradients at kink points), the entire LTM computation graph admits AD without approximation.
The key requirement is that all conditional logic (demand vs.\ supply binding, free-flow vs.\ congestion) must be expressed through $\min$/$\max$ rather than if-else branches, so that the computation graph remains connected regardless of the traffic regime.

An alternative to AD is the numerical differentiation method (e.g., finite-difference), which requires $O(|\tthh|)$ forward simulations per gradient.
This becomes impractical for high-dimensional parameter vectors such as network-scale parameters.
By contrast, reverse-mode AD computes the full gradient in a single backward pass regardless of $|\tthh|$.

The LTM has a further advantage for AD, namely, its coarse temporal discretization.
The CFL condition for the LTM is $\Dt \leq \min_l d_l / u_l$, where $d_l$ is the link length.
For a typical urban network with link lengths of 100--1000~m and free-flow speeds of 10--20~m/s, the maximum timestep is on the order of 5--100~s.
In contrast, cell-based simulation methods (e.g., CTM, METANET) need to employ shorter timestep and cell sizes to ensure high accuracy (e.g, 1--5~s).
Microscopic simulators typically use second or sub-second timesteps.
Since reverse-mode AD must back-propagate through each timestep sequentially, the computational and memory cost of gradient computation scales linearly with the number of timesteps $T_S$.
The LTM's large timestep means that a given simulation duration requires far fewer steps (e.g., $T_S = O(10^2)$--$O(10^3)$) compared to the CTM ($O(10^3)$--$O(10^5)$) or microscopic models ($O(10^5)$--$O(10^6)$), yielding proportionally faster and more memory-efficient gradient computation.

Furthermore, despite being a macroscopic model, the LTM allows the recovery of individual vehicle trajectories and travel times through its cumulative count representation and the virtual vehicle formulation, as explained in \cref{sec_virtual_vehicle}.
This means that trip-level objectives, such as the travel time of a specific OD pair departing at a given time (e.g., data from probe vehicles and connected vehicles), can also be differentiated with respect to the model parameters.
This bridges the gap between macroscopic simulation efficiency and microscopic output granularity.

\subsubsection{Differentiable LTM}\label{sec_diff_operators}

We describe two implementation choices needed to express the LTM in a form compatible with AD.

In the LTM formula, the demand and supply computations (\cref{eq_demand,eq_supply}) evaluate cumulative counts at fractional timestep indices $\tau = t + 1 - d_l/(u_l \Dt)$.
We use linear interpolation
\begin{align}
	N_{l,U}(\tau) = (1 - \delta) \cdot N_{l,U}(\lfloor \tau \rfloor) + \delta \cdot N_{l,U}(\lfloor \tau \rfloor + 1), \quad \delta = \tau - \lfloor \tau \rfloor,	\label{eq_interp}
\end{align}
which is differentiable with respect to both the stored cumulative count values and the index $\tau$ itself.
Since $\tau$ depends on the FD parameters $u_l$ and $w_l$, perturbations to these parameters alter the temporal offset at which past cumulative counts influence the current demand or supply, and the gradient correctly captures this coupling.

The virtual vehicle travel time (\cref{eq_exit_time,eq_chain_tt}) requires inverting the downstream cumulative count (i.e., finding the time $t$ at which $N_{l,D}(t) = N$ for a given $N$).
Since $N_{l,D}$ is stored as a discrete array and is monotonically non-decreasing, the inversion is implemented in two steps: first, the segment $[\lfloor t \rfloor, \lfloor t \rfloor + 1]$ containing the target value is identified by binary search; second, the fractional index within the segment is obtained by the same linear interpolation as \cref{eq_interp}, applied in reverse.
The segment selection is a discrete operation (non-differentiable), but the subsequent interpolation is differentiable with respect to the cumulative count values.
Combined with the $\max$ operation in \cref{eq_exit_time}, the resulting travel time function is piecewise differentiable with respect to the simulation parameters, enabling gradient-based optimization of trip-level objectives.
Furthermore, the time-dependent shortest path is automatically recovered from the results of the differentiable DUO.

\subsubsection{Differentiable INM}\label{sec_diff_inm}

The INM described in \cref{sec_node} is an iterative algorithm whose number of iterations depends on the current demand and supply values.
AD requires a static computation graph, so the variable-length iteration must be reformulated.
We express the INM as a fixed-length scan of $K = |\mathcal{L}_\nu^{\mathrm{in}}| + |\mathcal{L}_\nu^{\mathrm{out}}| + 1$ iterations, where each iteration is a closed-form update involving only arithmetic and $\min$/$\max$.

Consider node $\nu$ with inlinks $l = 1, \ldots, I$, outlinks $o = 1, \ldots, J$, turning fraction matrix $B = [b_{lo}]$, and merge priorities $\alpha_l$.
Let $\hat{q}_l^{(k)}$ and $\hat{q}_o^{(k)}$ denote the accumulated inflow and outflow allocations at iteration $k$, initialized as $\hat{q}_l^{(0)} = 0$ and $\hat{q}_o^{(0)} = 0$.
Each iteration proceeds as follows.

First, the {\it active set indicators} are computed.
An inlink $l$ is active if its allocation has not reached its demand and none of its target outlinks is saturated:
\begin{align}
	\delta_l^{\uparrow} = \mathbf{1}\!\big[\hat{q}_l^{(k)} < D_l\big] \cdot \prod_{o:\, b_{lo} > 0} \mathbf{1}\!\big[\hat{q}_o^{(k)} < S_o\big],	\label{eq_inm_active_in}
\end{align}
where $\mathbf{1}[\cdot]$ denotes the indicator function.
An outlink $o$ is active if it has remaining supply and at least one active inlink feeds it:
\begin{align}
	\delta_o^{\downarrow} = \mathbf{1}\!\big[\hat{q}_o^{(k)} < S_o\big] \cdot \mathbf{1}\!\big[\textstyle\sum_{l:\, b_{lo} > 0} \delta_l^{\uparrow} > 0\big].	\label{eq_inm_active_out}
\end{align}

Second, the {\it priority-weighted direction} is computed:
\begin{align}
	\phi_l = \delta_l^{\uparrow} \cdot \alpha_l, \qquad
	\phi_o = \sum_{l=1}^{I} b_{lo}\, \phi_l.	\label{eq_inm_direction}
\end{align}

Third, the {\it step size} $\theta$ is determined as the largest increment that does not violate any constraint:
\begin{align}
	\theta = \min\left\{
		\min_{l:\, \delta_l^{\uparrow}=1} \frac{D_l - \hat{q}_l^{(k)}}{\phi_l},~
		\min_{o:\, \delta_o^{\downarrow}=1} \frac{S_o - \hat{q}_o^{(k)}}{\phi_o}
	\right\}.	\label{eq_inm_theta}
\end{align}
At each iteration, $\theta$ drives at least one constraint to its bound, so the algorithm converges in at most $K = I + J + 1$ iterations.

Finally, the allocations are updated:
\begin{align}
	\hat{q}_l^{(k+1)} = \hat{q}_l^{(k)} + \theta\, \phi_l, \qquad
	\hat{q}_o^{(k+1)} = \hat{q}_o^{(k)} + \theta\, \phi_o.	\label{eq_inm_update}
\end{align}

After convergence (when no inlink is active), $\phi_l = 0$ for all $l$, and thus $\theta = 0$ regardless of the remaining capacity.
Subsequent iterations apply the identity $\hat{q}^{(k+1)} = \hat{q}^{(k)}$, contributing neither to the forward output nor to the backward gradient.
The procedure is summarized in \cref{alg_inm}.

\begingroup
\let\(\OrigLparen
\let\)\OrigRparen
\begin{algorithm}[t]
\caption{Fixed-length differentiable INM at node $\nu$}\label{alg_inm}
\begin{algorithmic}[1]
\Require Demands $D_l$, supplies $S_o$, turning fractions $B = [b_{lo}]$, priorities $\alpha_l$
\Ensure Inflow allocations $\hat{q}_l$, outflow allocations $\hat{q}_o$
\State $\hat{q}_l \gets 0$ for $l = 1, \ldots, I$; \quad $\hat{q}_o \gets 0$ for $o = 1, \ldots, J$
\For{$k = 1, \ldots, K$} \Comment{$K = I + J + 1$; fixed length}
	\State $\delta_l^{\uparrow} \gets \mathbf{1}[\hat{q}_l < D_l] \cdot \prod_{o:\, b_{lo}>0} \mathbf{1}[\hat{q}_o < S_o]$ for each $l$ \Comment{Active inlinks}
	\State $\delta_o^{\downarrow} \gets \mathbf{1}[\hat{q}_o < S_o] \cdot \mathbf{1}[\sum_{l:\, b_{lo}>0} \delta_l^{\uparrow} > 0]$ for each $o$ \Comment{Active outlinks}
	\State $\phi_l \gets \delta_l^{\uparrow} \cdot \alpha_l$; \quad $\phi_o \gets \sum_l b_{lo}\, \phi_l$ \Comment{Direction (\cref{eq_inm_direction})}
	\State $\theta \gets \min\big\{\min_{l:\,\phi_l > 0} (D_l - \hat{q}_l)/\phi_l,\;\min_{o:\,\phi_o > 0} (S_o - \hat{q}_o)/\phi_o\big\}$ \Comment{Step size (\cref{eq_inm_theta})}
	\State $\hat{q}_l \gets \hat{q}_l + \theta\, \phi_l$; \quad $\hat{q}_o \gets \hat{q}_o + \theta\, \phi_o$ \Comment{Update (\cref{eq_inm_update})}
\EndFor
\State \Return $\hat{q}_l$, $\hat{q}_o$
\end{algorithmic}
\end{algorithm}
\endgroup

The key property for AD is that each iteration (\cref{eq_inm_active_in}--\cref{eq_inm_update}) is expressed as arithmetic and $\min$/$\max$ operations on the continuous variables $D_l$, $S_o$, $\alpha_l$, and $b_{lo}$.
The indicator functions are implemented as comparisons that select between two continuous paths (e.g., $\phi_l = \alpha_l$ or $\phi_l = 0$), so the gradient with respect to $\alpha_l$ and $b_{lo}$ flows through the active constraints at each iteration.
The fixed iteration count $K$ ensures a static computation graph, and the natural zero-padding ($\theta = 0$ post-convergence) avoids any approximation.

\subsubsection{Differentiable DUO with logit model-based smoothing}\label{sec_diff_duo}

The DUO model (\cref{sec_route}) introduces three additional components that must be handled within the AD framework: instantaneous travel time computation, shortest-path-based routing, and handling of the zero gradient of any route choice pattern.

The instantaneous travel time (\cref{eq_tt_avg,eq_tt}) is computed by evaluating the cumulative curve $N(t,x)$.
The required operations are composed of subtraction (\cref{eq_tt_avg_k}), linear interpolation (\cref{eq_interp}), $\min$ (from Newell's formula), arithmetic, and $\min$/$\max$ (from the FD).
Consequently, the travel time $\tau_l(t)$ is differentiable with respect to all state variables and FD parameters by the same mechanisms described in \cref{sec_diff_operators}.

The Bellman--Ford algorithm produces a shortest path tree $\{\pi_\nu^s\}$ that is a discrete (integer-valued) function of the link travel times.
This discrete mapping is not differentiable in the classical sense.
However, the diverge ratios (\cref{eq_duo_beta}) that are actually used by the node models are {\it continuous} functions of the per-destination vehicle counts $\omega_\nu^s(t)$, which are themselves differentiable state variables.
A perturbation to $\tthh$ that does not change the shortest path tree leaves the route indicator $\mathbf{1}[\pi_\nu^s = o]$ unchanged; the gradient flows through the weights $\omega_\nu^s(t)$ in the numerator and denominator of \cref{eq_duo_beta}.
When a perturbation is large enough to change the shortest path tree, the diverge ratios shift discontinuously; however, this is a measure-zero event analogous to the $\min$/$\max$ kink points in the LTM.

In practice, gradient-based optimizers handle these discontinuities without difficulty, because the diverge ratios are piecewise smooth and the subgradients from the Bellman--Ford algorithm provide valid descent directions.

The per-destination outflow (\cref{eq_fifo}) is a ratio of differentiable cumulative counts.
Division by $N_{l,U}(t)$ is protected by a small constant to avoid division by zero; the gradient is well-defined whenever traffic is present on the link.

\begin{figure}[htp]
	\centering
%
%
%
%
%

	\includegraphics[clip, width=0.8\hsize]{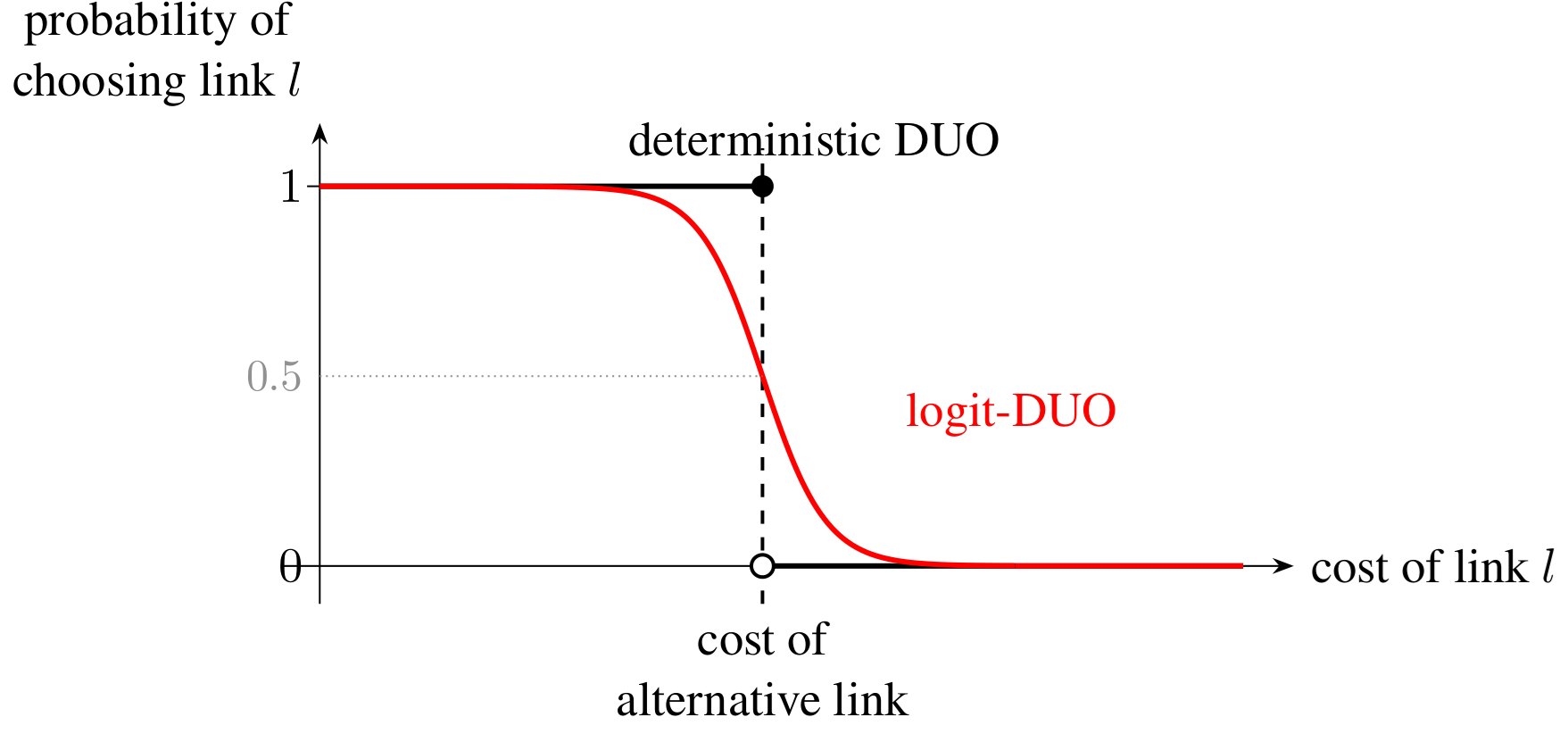}
	\caption{Conceptual illustration of logit-DUO.}
	\label{fig_logit_DUO}
\end{figure}

However, this deterministic DUO model has a significant limitation for particular applications.
That is, although this model is continuous and differentiable almost everywhere, the gradient of the shortest path with respect to link cost is {\it zero} almost everywhere (i.e., a step function).
This is because a small change in link cost usually does not change the path at all, and when the change is sufficiently large, the shortest path switches discontinuously.
In order to relax this limitation, we introduce a logit-based route choice model \citep{benakiva1999logit}.
For traffic with destination $s$ that arrives at node $\nu$ at time $t$, the turning probability to link $l$ is defined as
\begin{align}
	p(l, \nu; s, t) = \frac{\exp \(-\mu\mathrm{Cost}(l;s,t)\)}{\sum_{o \in {\cal L}_{\nu}^{\mathrm{out}}} \exp \(- \mu\mathrm{Cost}(o;s,t)\)},	\label{eq_logit_DUO}
\end{align}
where $\mu$ is the logit scale parameter and $\mathrm{Cost}(l;s,t)$ is the shortest path cost from link $l$ to destination $s$ at time $t$, which is simultaneously obtained by the aforementioned Bellman--Ford algorithm along with $\pi_\nu^s$.
\Cref{fig_logit_DUO} illustrates the relation between DUO and logit-DUO.
In fact, this can be considered a common approach to modeling route choice.
The total turning rate is then computed by aggregating $p(l, \nu; s, t)$ over destinations, weighted by the traffic volume bound for each destination.

In summary, the DUO and logit-DUO models preserve end-to-end differentiability: the gradient flows from the objective through the per-destination cumulative counts, through the diverge ratio computation, through the node models, and back to the FD and demand parameters.
Furthermore, the logit-DUO typically has a non-zero gradient with respect to link or path cost.
The use of physically grounded route choice models, rather than a differentiable surrogate model, ensures that the simulator produces traffic-theoretically consistent user optimum flows while maintaining gradient access.

\subsubsection{Simulation algorithm}\label{sec_algorithm}

The complete simulation procedure is summarized in this section.
The simulation state $\xx_t$ consists of the cumulative count arrays $\{N_{l,U}(t), N_{l,D}(t)\}_{l \in \mathcal{L}}$, origin demand queues $\{r_\nu(t)\}$, and absorbed vehicle counts at the destination nodes.
Each timestep applies the transition function $g$:
\begin{align}
	\xx_{t+1} = g(\xx_t, t; \tthh),	\label{eq_scan}
\end{align}
where $g$ is the composition of demand/supply computation, node model evaluation, and cumulative count update.

\begingroup
	\let\(\OrigLparen
	\let\)\OrigRparen
	\begin{algorithm}[t]
		\caption{Differentiable LTM: compilation and differentiation}\label{alg_sim}
		\begin{algorithmic}[1]
			\Require Parameter vector $\tthh$; network $\mathcal{G}$; scalar objective $J$ (as a function of state variables)
			\Ensure Gradient $\nabla_{\tthh} J$ or Jacobian-vector product
			\Statex \textbf{--- Compilation (once) ---}
			\State Define one-step transition $g(\xx_t, t; \tthh)$: \Comment{Piecewise differentiable}
			\Statex \quad Demand/supply via $\min$/$\max$ (\cref{eq_demand,eq_supply}) with linear interpolation (\cref{eq_interp})
			\Statex \quad Travel time $\tau_l$ from $N$-curves (\cref{eq_tt}); shortest paths via Bellman--Ford;
			\Statex \quad \quad logit-DUO diverge ratios $\beta_{\nu o}$ from per-destination weights (\cref{eq_duo_beta,eq_logit_DUO})
			\Statex \quad Node models via $\min$/$\max$ fixed-$K$ INM with $\min$/$\max$ (\cref{alg_inm})
			\Statex \quad Per-destination flow split by FIFO (\cref{eq_fifo})
			\Statex \quad Cumulative count update (\cref{eq_update_U,eq_update_D}) and queue update
			\State Define $F(\tthh) = \textsc{Scan}(g, \xx_0, [0, \ldots, T_S{-}1]; \tthh)$ \Comment{Forward simulation}
			\State Define $L(\tthh) = J(F(\tthh); \tthh)$ \Comment{Scalar loss}
			\State JIT-compile $L$ \Comment{Trace computation graph; fuse operations}
			\Statex \textbf{--- Differentiation mode (per call) ---}
			\Statex \textit{Option A: Reverse mode} (for $|\tthh| \gg 1$, scalar $J$) \Comment{$O(T_S)$ time, $O(\sqrt{T_S} n_x)$ memory}
			\State $\nabla_{\tthh} J \gets \textsc{Grad}(L)(\tthh)$ \Comment{Backprop through \textsc{Scan} with checkpointing}
			\Statex \textit{Option B: Forward mode} (for memory-less per-parameter sensitivity) \Comment{$O(T_S)$ time per direction}
			\State $\del J / \del \theta_i \gets \textsc{Jvp}(L, \tthh, \mathbf{e}_i)$ \Comment{Propagate unit tangent $\mathbf{e}_i$ through \textsc{Scan}}
		\end{algorithmic}
	\end{algorithm}
\endgroup

\cref{alg_sim} summarizes the differentiation pipeline.
The one-step transition $g$ encapsulates the LTM computations (\cref{sec_ltm,sec_node}), expressed through arithmetic and $\min$/$\max$ so that the composition is piecewise differentiable.
In the pseudocode, $\textsc{Scan}$ denotes a sequential fold that iteratively applies $g$ over all timesteps starting from the initial state $\xx_0$ (i.e., cumulative counts, queues, and absorbed counts at $t = 0$, typically all zeros).
$\textsc{Grad}$ denotes the reverse-mode gradient operator, $\textsc{Jvp}$ denotes the forward-mode Jacobian-vector product, $\mathbf{e}_i$ is the $i$-th standard basis vector in parameter space, and JIT refers to just-in-time compilation that traces and fuses the computation graph.
These correspond to the JAX primitives \texttt{jax.lax.scan}, \texttt{jax.grad}, \texttt{jax.jvp}, and \texttt{jax.jit}, respectively.

Since the entire computation graph is differentiable, both reverse-mode and forward-mode AD are applicable.
Reverse-mode AD computes the full gradient $\nabla_{\tthh} J$ in a single backward pass by backpropagation through the \textsc{Scan}, which is mathematically equivalent to the discrete adjoint method \citep{Pontryagin1962adjoint} but derived automatically.
This mode is efficient when the objective is scalar and $|\tthh|$ is large (e.g., OD demand calibration), as the cost is independent of the parameter dimension.

Forward-mode AD, conversely, propagates a tangent vector through the \textsc{Scan} in a single forward pass, computing the directional derivative $\del J / \del \theta_i$ for one parameter at a time.
This mode is efficient when $|\tthh|$ is small or when the Jacobian-vector product (rather than the full gradient) is needed, as it requires no additional memory beyond the forward simulation state.
Higher-order derivatives (e.g., Hessian-vector products) are also available by composing the two modes.

Two additional technical strategies are implemented to enhance computational efficiency.
First, the node model evaluation within each timestep is vectorized across all nodes: the transfer logic for each node type is computed simultaneously via batched array operations, and the appropriate result for each node is selected through element-wise conditionals, enabling data-parallel execution on accelerator hardware such as Graphics Processing Units (GPUs).
Second, the \textsc{Scan} carry state is restricted to a sliding window of recent cumulative counts whose width is determined by the maximum free-flow travel time and is thus sufficient for all demand and supply lookups, while per-step incremental flows are emitted as auxiliary scan outputs.
The full cumulative count trajectories are then reconstructed by summation after the scan completes, reducing the per-step carry dimension and the associated memory overhead of reverse-mode differentiation.
These strategies allow us to simulate large-scale scenarios very quickly using a GPU (see \cref{sec_chicago} for an example).

\subsubsection{Differentiable parameters and computational cost}\label{sec_diff_scope}

The differentiable parameters span all physically meaningful inputs:
FD parameters ($u_l, \kappa_l, w_l$), OD demand ($q_{rs}(t)$), and merge priorities ($\alpha_l$).
In the fixed turning rate mode, diverge ratios ($\beta_{\nu o}$) and turning fractions ($b_{lo}$) are additionally differentiable.
The cost of the backward pass is proportional to that of the forward simulation (typically 2--5 times) and is independent of the parameter dimension $|\tthh|$.

\subsection{Summary}\label{sec_method_summary}

The proposed simulator, named {\it UNsim}, combines the LTM for link dynamics, a suite of node models (origin/destination, dummy, diverge, merge, and the INM for general nodes), and an end-to-end differentiable formulation.
The simulation computes the $N$-curve (cumulative vehicle count) as the primary state variable, as opposed to the vehicle trajectory $X$ used by Lagrangian (agent-based) simulators such as {\it UXsim} \citep{seo2025joss}, hence the name.
The entire mapping from parameters to objective functions is differentiable almost everywhere without significant approximations, enabling exact gradient computation for calibration, sensitivity analysis, and optimization tasks.
Furthermore, the end-to-end differentiable implementation provides a natural foundation for integrating neural network modules with the physical simulator, potentially enabling physics-informed hybrid models in future work.
The computation code is implemented using Python and JAX (with GPU-acceleration), and released as an open-source Python package.

\section{Numerical examples}\label{sec_numerical}

Two numerical examples are presented.
First, in \cref{sec_toy}, we investigate whether the values of AD-based partial derivatives of the proposed model are reasonable or not.
We use a small toy network for interpretable analysis.
Second, in \cref{sec_chicago}, we apply the proposed model to a dynamic congestion pricing optimization problem in a large-scale, realistic network called the Chicago-Sketch dataset and solve a differentiable simulation-based traffic optimization problem.

\subsection{Verification of partial derivatives in simple toy networks}\label{sec_toy}


In order to verify that the partial derivatives computed by the differentiable simulator are qualitatively and quantitatively reasonable, we use a simple Y-shaped merge network and a two-routes network.

\subsubsection{Merge network}

\begin{figure}[htp]
	\centering
	\includegraphics[clip, width=0.6\hsize]{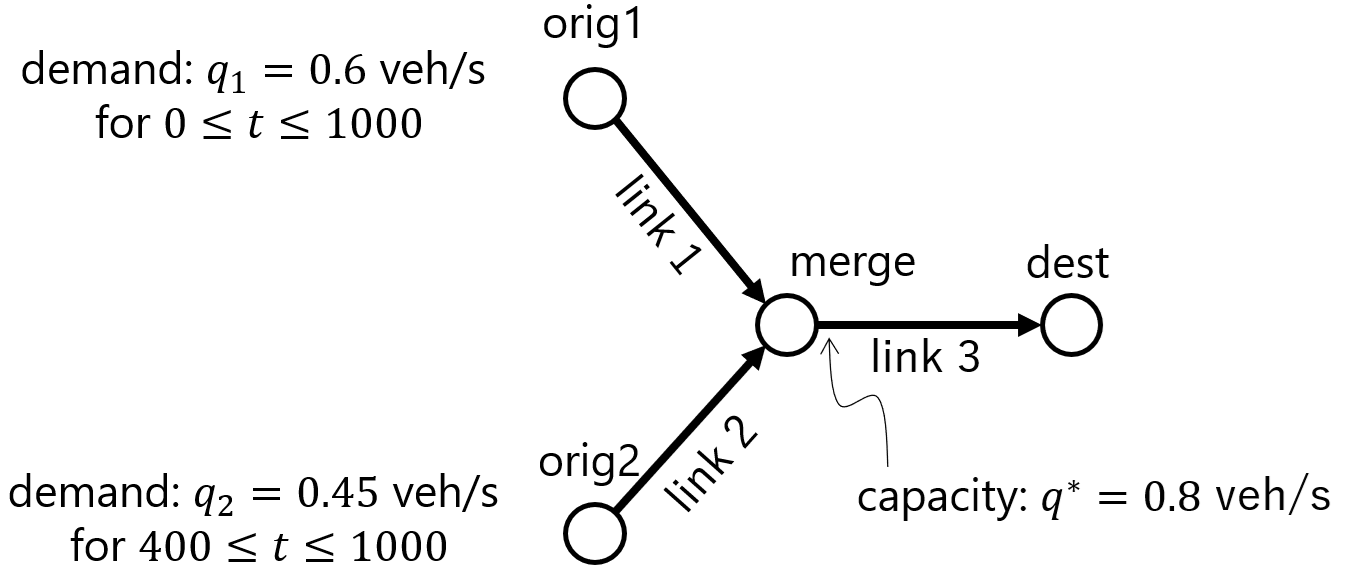}
	\caption{Merge network}
	\label{fig_merge_network}
\end{figure}

\begin{figure}[htp]
	\centering
	\subfloat[$t = 300$ s]{\includegraphics[clip, width=0.4\hsize]{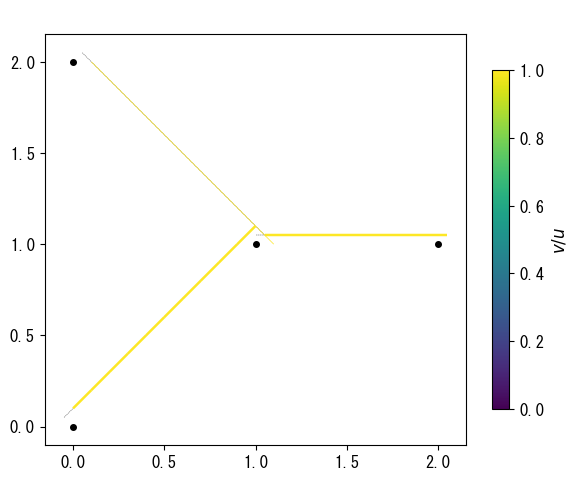}\label{net_t300}}~
	\subfloat[$t = 800$ s]{\includegraphics[clip, width=0.4\hsize]{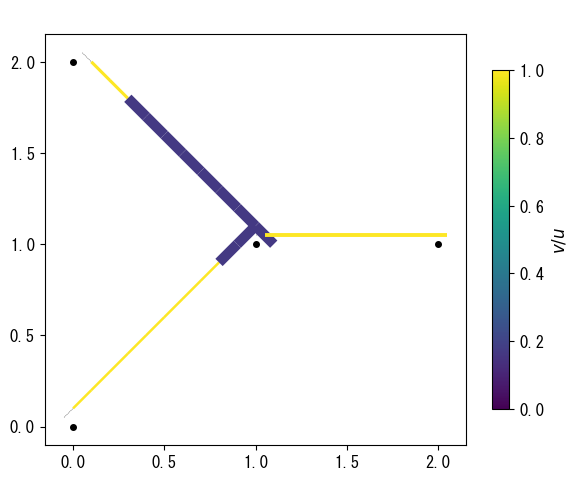}\label{net_t800}}
	\caption{Network speed (normalized by free-flow speed). The width represents the density at the segment.}
	\label{merge_network_speed}
\end{figure}

\begin{figure}[htp]
	\centering
	\subfloat[Link 1]{\includegraphics[clip, width=0.6\hsize]{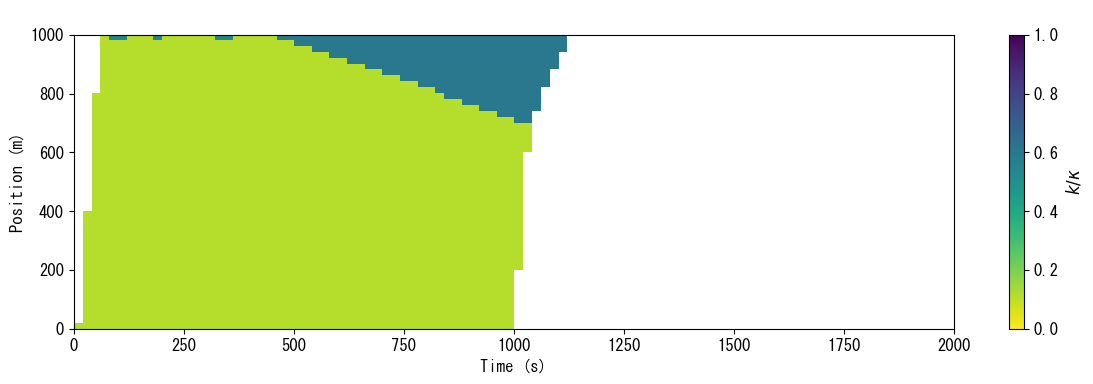}\label{tsd_link1}}\\
	\subfloat[Link 2]{\includegraphics[clip, width=0.6\hsize]{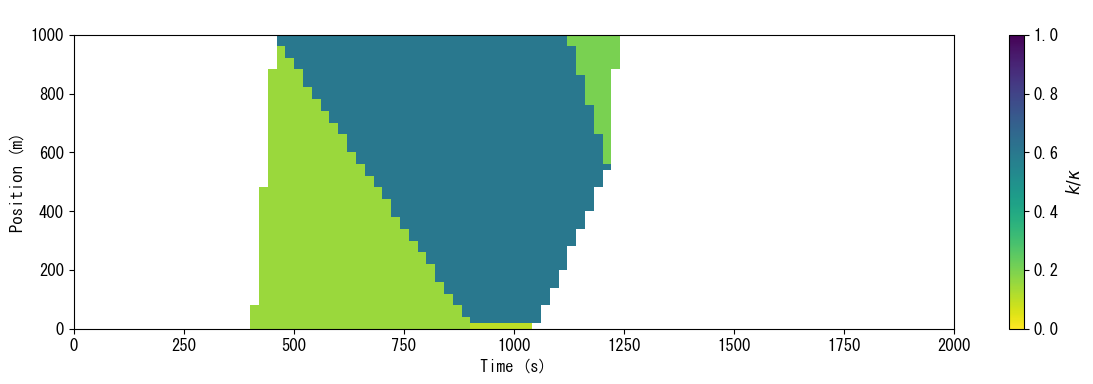}\label{tsd_link2}}\\
	\subfloat[Link 3]{\includegraphics[clip, width=0.6\hsize]{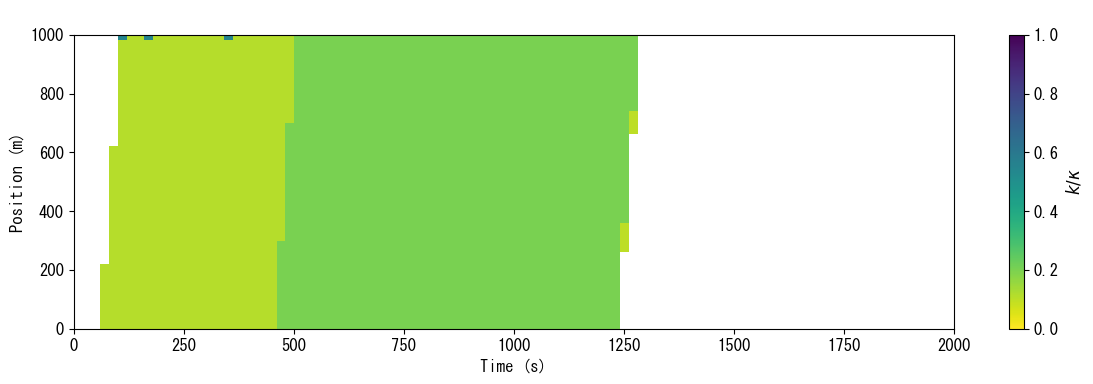}\label{tsd_link3}}
	\caption{Time-space diagrams of density (normalized by jam density). Note that these values are obtained by numerically differentiating $N$ for visualization purposes, and thus some numerical noises exist, especially near the link borders.}
	\label{fig_tsds}
\end{figure}

We consider a merge network consisting of four nodes (two origins, one merge, one destination) and three links, as illustrated in \cref{fig_merge_network}.
All links have identical FD parameters and length: $u = 20$~m/s, $q^*=0.8$~veh/s, $\kappa = 0.2$~veh/m, and $d = 1000$~m.
The merge priority weights are also equal for both incoming links ($\alpha_1 = \alpha_2 = 1$).
Origin~1 generates demand at a rate of $q_1 = 0.45$~veh/s during $t \in [0, 1000]$~s, and origin~2 generates demand at $q_2 = 0.6$~veh/s during $t \in [400, 1000]$~s.
The combined demand ($0.45 + 0.6 = 1.05$~veh/s) exceeds the downstream link capacity ($0.8$~veh/s) during the overlap period, causing congestion at the merge.
The simulation uses a timestep of $\Dt = 5$~s and a duration of $T_{\max} = 2000$~s.
Forward simulation results are visualized in \cref{merge_network_speed,fig_tsds}; merging congestion consistent with the kinematic wave theory was accurately simulated.


First, we define the total travel time (TTT) as an objective function:
\begin{align}
	\mathrm{TTT} = \sum_{l \in \mathcal{L}} \sum_{t=0}^{T_S-1} \max\{N_{l,U}(t) - N_{l,D}(t),~ 0\} \cdot \Dt + \sum_{\nu \in \mathcal{N}_{\mathrm{orig}}} \sum_{t=0}^{T_S-1} r_\nu(t) \cdot \Dt,	\label{eq_ttt2}
\end{align}
where $\mathcal{N}_{\mathrm{orig}}$ denotes the set of origin nodes.
The first term accounts for vehicles on links, and the second term accounts for vehicles in vertical queues at origins.

Once the forward simulation with a given parameter vector $\tthh$ is completed, the partial derivative of $\mathrm{TTT}$ with respect to any element of $\tthh$ is obtained by a single backward pass.
For example, the partial derivatives of TTT with respect to the demand rates are computed as
\begin{align}
	\frac{\del \mathrm{TTT}}{\del q_1} = 437455.688, \qquad
	\frac{\del \mathrm{TTT}}{\del q_2} = 421503.938,
\end{align}
where $q_1$ and $q_2$ denote the demand rates at origin~1 and origin~2, respectively.
Both values are positive, indicating that an increase in demand from either origin increases the TTT---a reasonable result.
It would be worth emphasizing again that this is not a numerical differentiation, a finite difference, or any other approximation; they are simultaneously computed by one AD operation.

With respect to the FD parameters, the following values are obtained:
\begin{align}
	\frac{\del \mathrm{TTT}}{\del u_1} = -1278.282, \qquad
	\frac{\del \mathrm{TTT}}{\del u_2} = -616.873, \qquad
	\frac{\del \mathrm{TTT}}{\del u_3} = -2024.685.
\end{align}
All values are negative, meaning that increasing the free-flow speed of each link decreases TTT.
Furthermore, the magnitude is largest for link~3, which is used by all traffic after the merge.

Now let us look at individual link states:
\begin{align}
	\frac{\del \mathrm{TTTlink1}}{\del \alpha_1} = -45900.031, \qquad
	\frac{\del \mathrm{TTTlink2}}{\del \alpha_1} = 40725.027, 
\end{align}
where $\mathrm{TTTlink}l$ denotes TTT on link $l$ only, defined similarly to \cref{eq_ttt2}.
They mean that TTT on link 1 decreases if the merging priority of link 1 increases; conversely, TTT on link 2 increases if the merging priority of link 1 increases.

We now examine some individual vehicle trajectories.
Let $\mathrm{TT}(t, r, s)$ be the travel time of a vehicle that departs from origin $r$ at time $t$ and travels to destination $s$.
As explained in \cref{sec_virtual_vehicle}, this value can be directly computed from the LTM by using the cumulative count.
We obtain the following values:
\begin{align}
	&\frac{\del \mathrm{TT}(100, \mathrm{orig1}, \mathrm{dest})}{\del \alpha_1} = 0.000, \qquad
	\frac{\del \mathrm{TT}(500, \mathrm{orig1}, \mathrm{dest})}{\del \alpha_1} = -56.250, \\
	&\frac{\del \mathrm{TT}(100, \mathrm{orig2}, \mathrm{dest})}{\del \alpha_1} = 0.000, \qquad
	\frac{\del \mathrm{TT}(500, \mathrm{orig2}, \mathrm{dest})}{\del \alpha_1} = 75.000.
\end{align}
Around $t=100$~s, there is no congestion.
Therefore, the merging priority ratio does not have any impact on vehicles.
Around $t=500$~s, congestion exists due to the merge.
Therefore, increasing the merging priority ratio of link 1 decreases the travel time of vehicles departing from origin 1, and increases that of those departing from origin 2.

In order to quantitatively verify the AD of the proposed model, we compare it with values obtained by the finite difference method.
Note that finite difference is a numerical approximation and is not necessarily accurate, especially near non-differentiable points. 
Furthermore, the relationship between the magnitude of perturbation size $\epsilon$ and the accuracy of the numerical derivative is not clear a priori; thus, we use the central differentiation shown in \cref{eq_central} with $\epsilon$ ranging from $10^{-5}$ to $10^{-1}$.
Nevertheless, the finite difference method could be useful as a baseline.
\Cref{table_ad_vs_finite_merge} summarizes the results.
The AD values generally agree with those of the finite difference method, which supports the accuracy of the proposed model.

\begin{table}[htbp]
	\centering
	\caption{Comparison between AD and finite difference in the merge network.}
	\label{table_ad_vs_finite_merge}
	\scriptsize
	\begin{tabular}{lrrrrrr}
	    \toprule
	    \multirow{2}{*}{\textbf{Partial derivative}} & \multirow{2}{*}{\textbf{AD}} & \multicolumn{5}{c}{\textbf{Finite Difference}} \\
	    \cmidrule(lr){3-7}
	    & & $\epsilon=10^{-1}$ & $\epsilon=10^{-2}$ & $\epsilon=10^{-3}$ & $\epsilon=10^{-4}$ & $\epsilon=10^{-5}$ \\
	    \midrule
		$\partial$TTT / $\partial q_1$                        & $437455.688$ & $459549.844$ & $449726.563$ & $448187.500$ & $455546.875$ & $346875.000$ \\
		$\partial$TTT / $\partial q_2$                        & $421503.938$ & $439550.078$ & $434210.938$ & $434820.313$ & $437031.250$ & $401562.500$ \\
		$\partial$TTT / $\partial u_1$                             & $-1278.282$  & $-1335.547$  & $-1337.500$  & $-1320.313$  & $-1718.750$  & $1562.500$   \\
		$\partial$TTT / $\partial u_2$                             & $-616.873$   & $-589.141$   & $-589.844$   & $-578.125$   & $-468.750$   & $-2343.750$  \\
		$\partial$TTT / $\partial u_3$                             & $-2024.685$  & $-2025.547$  & $-2025.781$  & $-2023.438$  & $-1406.250$  & $0.000$      \\
		$\partial$TTTlink1 / $\partial \alpha_1$                      & $-45900.031$ & $-46024.531$ & $-45603.516$ & $-44746.094$ & $-50820.313$ & $-110546.875$ \\
		$\partial$TTTlink2 / $\partial \alpha_1$                      & $40725.027$  & $36414.277$  & $35819.727$  & $35679.688$  & $34003.906$  & $22851.563$  \\
		$\partial$TT(100, orig1, dest) / $\partial \alpha_1$    & $0.000$      & $0.000$      & $0.000$      & $0.000$      & $0.000$      & $0.000$      \\
		$\partial$TT(500, orig1, dest) / $\partial \alpha_1$    & $-56.250$    & $-56.818$    & $-56.256$    & $-56.244$    & $-56.458$    & $-57.983$    \\
		$\partial$TT(100, orig2, dest) / $\partial \alpha_1$    & $0.000$      & $0.000$      & $0.000$      & $0.000$      & $0.000$      & $0.000$      \\
		$\partial$TT(500, orig2, dest) / $\partial \alpha_1$    & $75.000$     & $75.000$     & $75.000$     & $75.043$     & $74.463$     & $73.242$     \\
		\bottomrule
	\end{tabular}
\end{table}

\subsubsection{Two-routes network}

\begin{figure}[htp]
	\centering
	\includegraphics[clip, width=0.99\hsize]{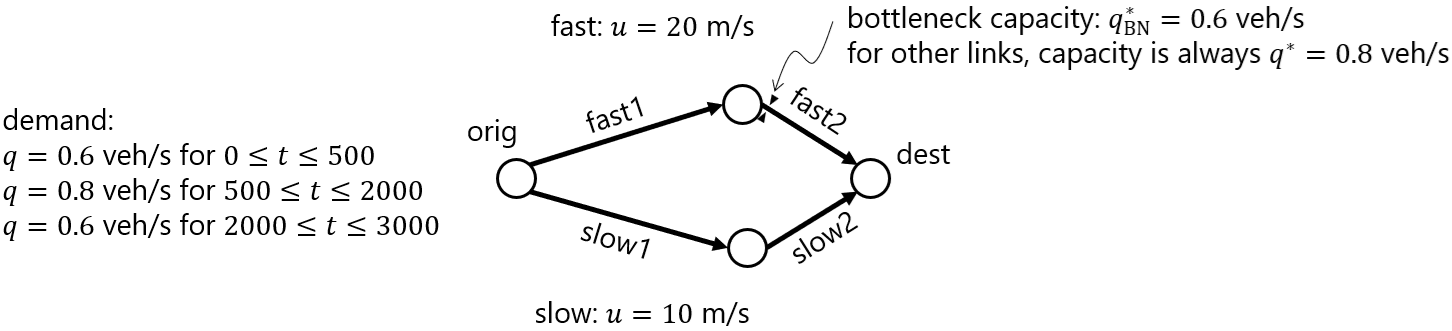}
	\caption{Two-routes network}
	\label{route_choice_network}
\end{figure}

\begin{figure}[htp]
	\centering
	\subfloat[Fast route. The bottleneck is at 1000 m location.]{\includegraphics[clip, width=0.6\hsize]{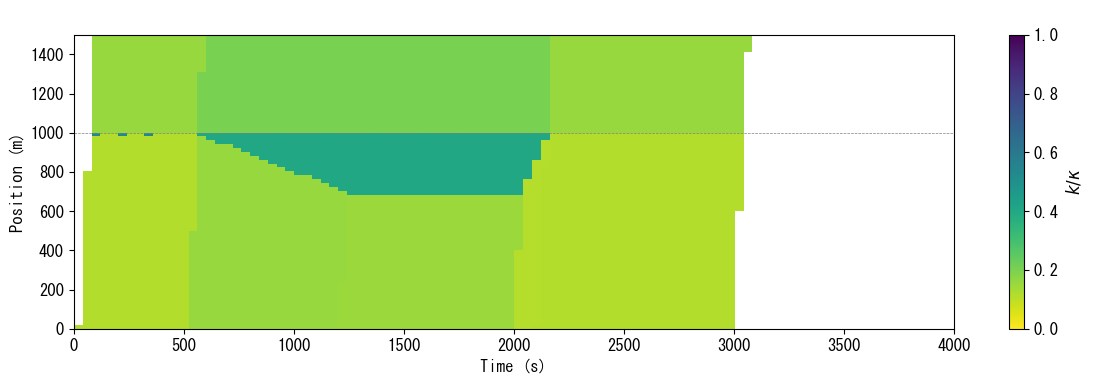}\label{tsd_fast}}\\
	\subfloat[Slow route]{\includegraphics[clip, width=0.6\hsize]{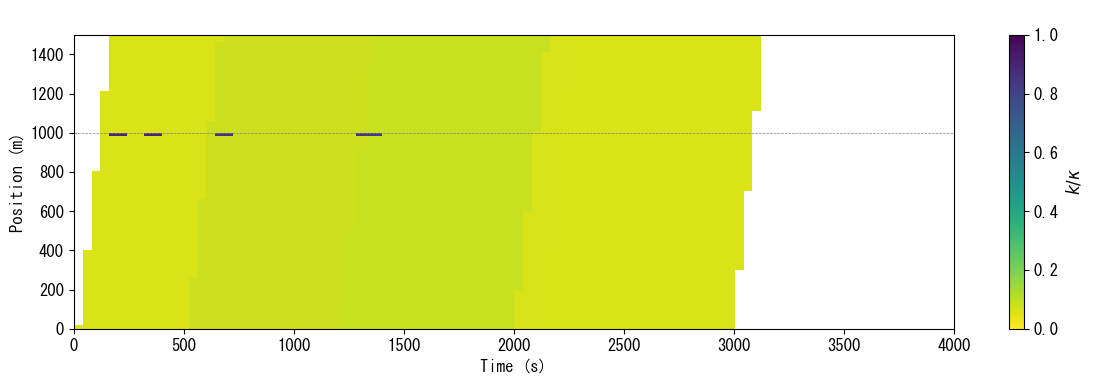}\label{tsd_slow}}
	\caption{Time--space diagrams of normalized density in the two-route network.}
	\label{tsd_route_choice}
\end{figure}

Now we consider a network with single OD pair and two routes shown in \cref{route_choice_network}.
There are two routes: a {\it fast route} with a bottleneck of capacity $q_{\mathrm{BN}}^*$ and a {\it slow route} with no bottleneck.
The demand is set as follows: the flow is lower than the bottleneck capacity at the beginning, higher than the capacity in the middle, and back to a low value at the end.
For the detailed specification, see \cref{route_choice_network}.
The forward simulation result is illustrated in \cref{tsd_route_choice}.
It can be seen that, during the high demand period, traffic congestion occurred on the fast route and the flow on the slow route increased.
Once the traffic reached equilibrium (around $t=1200$~s), it remained in the steady state until the high demand period ended.
Note that, since the logit model was used, a small flow was observed on the slow route as well during the low demand period although the travel time was longer than that on the fast route.
We conclude that the logit-DUO route choice model behaved reasonably by responding to the time-varying traffic situation endogenously.

We compute partial derivatives with respect to the bottleneck capacity $q_{\mathrm{BN}}^*$.
The results are summarized in \cref{table_ad_vs_finite_routechoice} along with those approximated by the finite difference method, where ATT$l$ denotes the average travel time of link $l$, and $\mathrm{TT}(t, p)$ denotes the travel time of a vehicle that starts traveling path $p$ at time $t$.
Qualitatively, the AD values are reasonable.
For example, the partial derivatives of the following variables are negative: $\mathrm{TTT}$, $\mathrm{ATT}$ of the link upstream of the bottleneck, and the travel time of vehicles on the fast route during the congested period.
Furthermore, those of the following variables are almost zero, meaning that they are not affected by the bottleneck capacity: the average travel time of links downstream of the bottleneck or on the other route, and the travel time of vehicles that did not experience congestion.
Quantitatively, the AD values generally agree with those of the finite difference method.

\begin{table}[htbp]
	\centering
	\caption{Comparison between AD and finite difference in the two-route network.}
	\label{table_ad_vs_finite_routechoice}
	\scriptsize
	\begin{tabular}{lrrrrrr}
		\toprule
	    \textbf{Partial derivative} & \multirow{2}{*}{\textbf{AD}} & \multicolumn{5}{c}{\textbf{Finite Difference}} \\
	    \cmidrule(lr){3-7}
	    {\bf ~with respect to $q^*_{\mathrm{BN}}$} & & $\epsilon=10^{-1}$ & $\epsilon=10^{-2}$ & $\epsilon=10^{-3}$ & $\epsilon=10^{-4}$ & $\epsilon=10^{-5}$ \\
	    \midrule
		TTT                          & $-586313.750$ & $-373412.344$ & $-608487.500$ & $-585906.250$ & $-613515.625$ & $-507031.250$ \\
		ATTfast1                    & $-314.249$    & $-213.364$    & $-327.942$    & $-312.449$    & $-361.748$    & $-220.490$    \\
		ATTfast2                    & $-0.001$      & $-0.000$      & $0.072$       & $-1.287$      & $31.567$      & $-41.103$     \\
		ATTslow1                    & $0.010$       & $-0.027$      & $-0.154$      & $-0.328$      & $-9.422$      & $-13.733$     \\
		ATTslow2                    & $-0.005$      & $-0.000$      & $0.002$       & $0.000$       & $0.114$       & $5.341$       \\
		TT(100, fast)               & $0.000$       & $0.000$       & $0.000$       & $0.000$       & $0.000$       & $0.000$       \\
		TT(1500, fast)              & $-570.359$    & $-390.270$    & $-608.917$    & $-570.251$    & $-576.782$    & $-439.453$    \\
		TT(100, slow)               & $0.000$       & $0.000$       & $0.000$       & $0.000$       & $0.000$       & $0.000$       \\
		TT(1500, slow)              & $0.001$       & $0.000$       & $0.000$       & $-0.061$      & $0.000$       & $0.000$       \\
		\bottomrule
	\end{tabular}
\end{table}

\subsection{Dynamic congestion pricing optimization in Chicago-Sketch network}\label{sec_chicago}

Now, we apply the proposed model to a dynamic congestion pricing optimization problem in a large network to validate the model's overall capability.
This is a suitable problem to investigate the unique features of the model: endogenous dynamic route choice and scalability for large scenarios.


\subsubsection{Optimization problem formulation}

Consider a situation where a time-varying toll is charged on each link in a network.
The toll on link $l$ is determined by a step function:
\begin{align}
	\cT_l(t) = \cT_l^i	\qquad t \in [i\Delta t_{\textrm{toll}}, (i+1)\Delta t_{\textrm{toll}}), \forall i,
\end{align}
where $\cT_l^i$ is a constant toll for time period $i$, and $\Delta t_{\textrm{toll}}$ is the period duration.
Our problem is to determine the values of $\cT_l^i$ for all links $l$ and time periods $i$.

The optimization problem is defined as
\begin{align}
	&\min_{\cT_l^i, \forall l,i} J = \mathrm{TTT} + \lambda \sum_{l,i} \(\cT_l^i\)^2,	\label{eq_toll_obj}\\
	&\mathrm{s.t.~} \cT_l^i \geq 0, \qquad \forall l,i,
\end{align}
where the first term in the objective function is the total travel time defined by \cref{eq_ttt2}, and the second term is an L2 regularization term to prevent excessively large tolls and the potentially negative effects of tolls that do not have any impact on traffic (e.g., tolls charged on links with no traffic).
Parameter $\lambda = 0.001$ is a small weight parameter for L2 regularization.

For optimization, we employ {\it Adam} \citep{Kingma2015adam}, one of the standard gradient-based optimization algorithms.
The gradient $\nabla_{\bm{\cT}} J$ is computed by reverse-mode AD through the entire DUO simulation, including the dynamic shortest-path routing and per-destination flow tracking.
We also use gradient clipping that shrinks the gradient vector if its L2 norm exceeds $2 \times 10^6$ and a projection $\cT_l^i \geq 0$ for the non-negative constraints.
The other hyperparameters are as follows: learning rate $\eta = 7.0$, momentum parameters $\beta_1 = 0.9$ and $\beta_2 = 0.999$.

\subsubsection{Network and simulation setting}\label{sec_chicago_simsetting}

\begin{figure}[htp]
	\centering
	\includegraphics[clip, width=0.8\hsize]{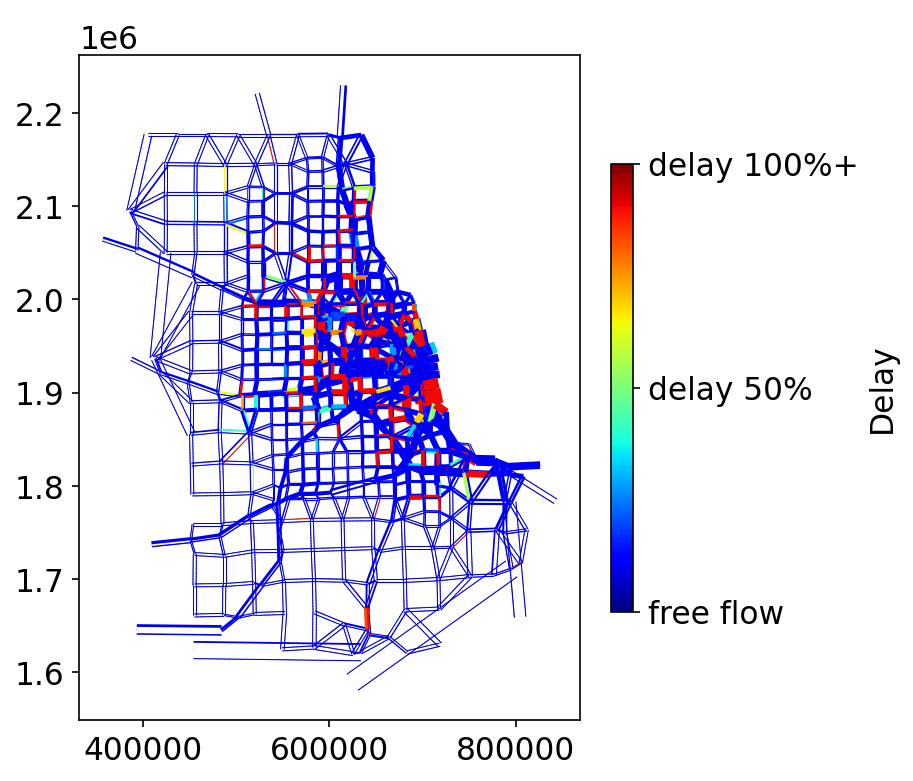}
	\caption{Simulated average link delay in the Chicago-Sketch data scenario without pricing. ``Delay'' is the ratio of the excess of the average link travel time over the free-flow travel time.}
	\label{toll_network_avg_baseline}
\end{figure}

We use the Chicago-Sketch dataset published by \citet{tnrct2021network}, which is a common dataset for this kind of analysis.
It contains 927 nodes, \num{2557} links, \num{17963} OD pairs, and approximately 1 million vehicles; note that the network data was slightly adjusted in order to fit the technical requirements of the proposed model.
The demand data was also modified to make it suitable for a dynamic pricing problem with a 3-hour simulation duration.
The original demand pattern was assumed to continue for 1 hour.
Then, the time periods were equally divided into three, and the flow rate in the second period was multiplied by 1.5, mimicking a peak period.
The total number of vehicles was \num{1165433}.
The number of decision variables ($\cT_l^i$) was \num{15320} (383 congested links $\times$ 40 time periods).
\cref{toll_network_avg_baseline} shows the simulated results without pricing.
Severe downtown congestion is observed.

Model parameter settings are as follows.
The LTM's timestep size $\Dt$ was 33.3 sec.
The DUO's route choice interval $\Dt_{\mathrm{route}}$ and pricing update time $\Delta t_{\textrm{toll}}$ were 300 sec.
We use logit-DUO with scale parameter $\mu=1/10$ (i.e., near-deterministic choice).
For the link FD and demand settings, we mostly adopt the configuration of the original dataset; readers are referred to \url{https://github.com/toruseo/UNsim} for the specifications as they are too detailed to describe here.
This simulation scenario was converted into a computational graph for AD with the following size: 706 nodes and 1992 edges for one time step, corresponding to 446k nodes and 1.41M edges in total.

For computation, a cloud server with a GPU was used.
The detailed specification is as follows: 1x NVIDIA GH200 (VRAM 96GB), 64x vCPUs (ARM Neoverse-V2 2.0GHz), 432 GB RAM.
Thanks to GPU-accelerated parallel computation implemented in the proposed model, the simulation speed itself is very fast.
One evaluation took around 0.8 sec on average: 0.2 sec for forward simulation and 0.6 sec for reverse-AD.

\subsubsection{Results}

The total number of iterations was set to \num{10000} without a convergence check.
The computation time was \num{8373} seconds, corresponding to 0.8 seconds per iteration.
The TTT without pricing was \num{1120100} veh-hr, and that with optimal pricing was \num{509743} veh-hr, representing a 54.5{\%} reduction.
The average toll per link was about 5 minutes of travel time, which can be roughly converted to a monetary value of 2 USD.

\begin{figure}[htp]
	\centering
	\includegraphics[clip, width=0.8\hsize]{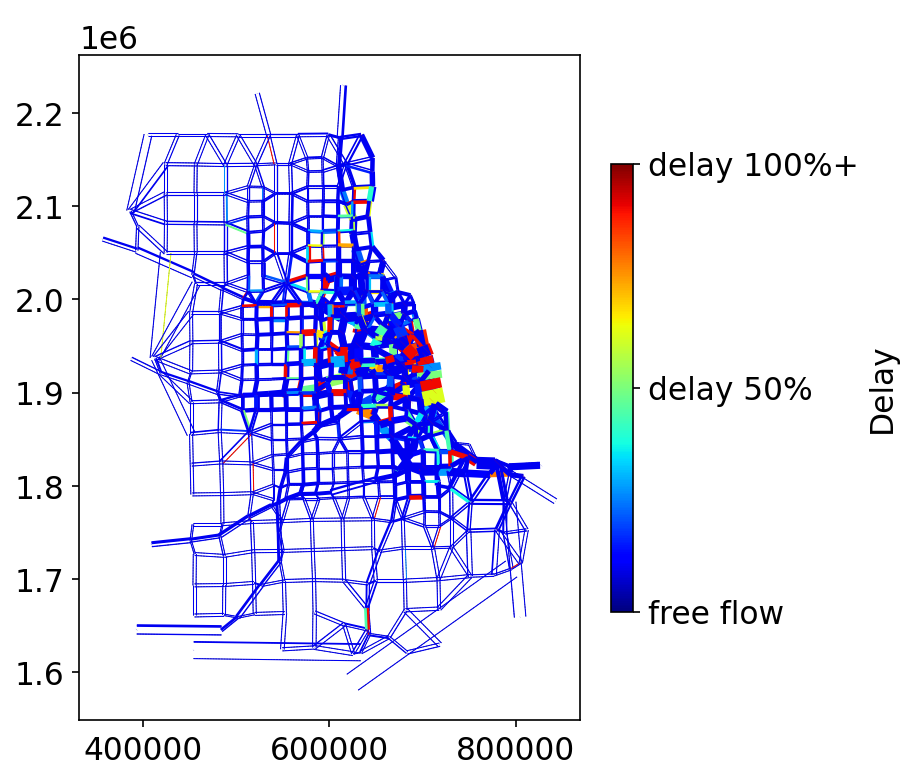}
	\caption{Average link delay with the best pricing.}
	\label{toll_network_avg_tolled}
\end{figure}

\Cref{toll_network_avg_tolled} shows the average network state in the best pricing case.
By comparing with the no-toll case (\cref{toll_network_avg_baseline}), we can see that severe congestion in downtown was mitigated significantly.

\begin{figure}[htp]
	\centering
	\includegraphics[clip, width=0.6\hsize]{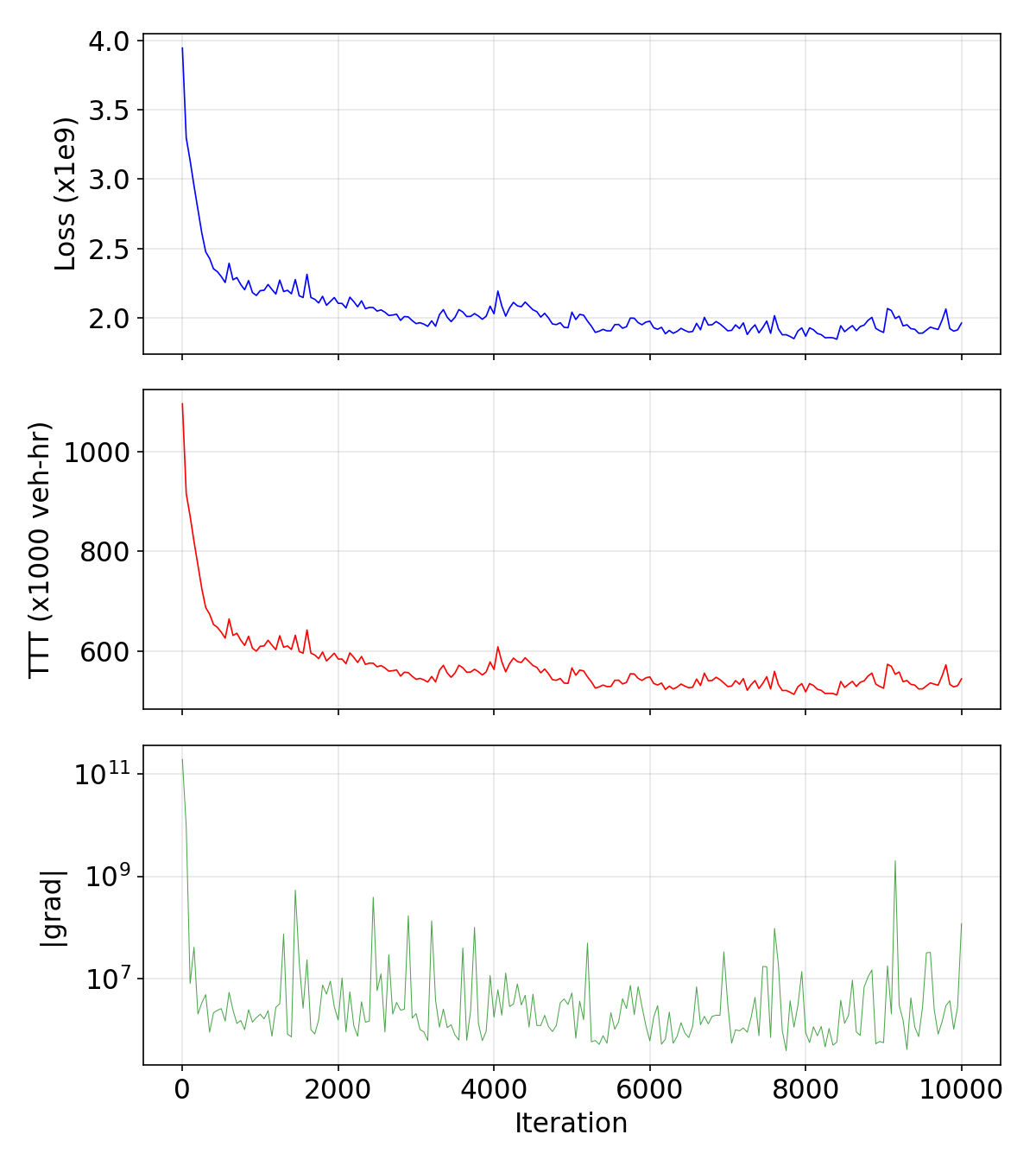}
	\caption{Convergence of the objective function, TTT, and gradient.}
	\label{toll_convergence}
\end{figure}

\Cref{toll_convergence} summarizes the convergence process over the iteration.
It can be seen that the objective function (with the L2 regularization term) and TTT decreased stably and almost converged around the 3000th iteration.
The norm of the gradient behaved reasonably: the gradient was very large at the beginning, and then it mostly stayed in some range ($10^6$--$10^7$) after the initial stage.
No significant instabilities were observed, thanks to the exact differentiation.
This convergence process could be further improved by implementing iteration termination criteria or fine-tuning the optimizer (e.g., adaptive adjustment of the step size), but such technical improvements are beyond the scope of this work.

\begin{figure}[htp]
	\centering
	\includegraphics[clip, width=0.8\hsize]{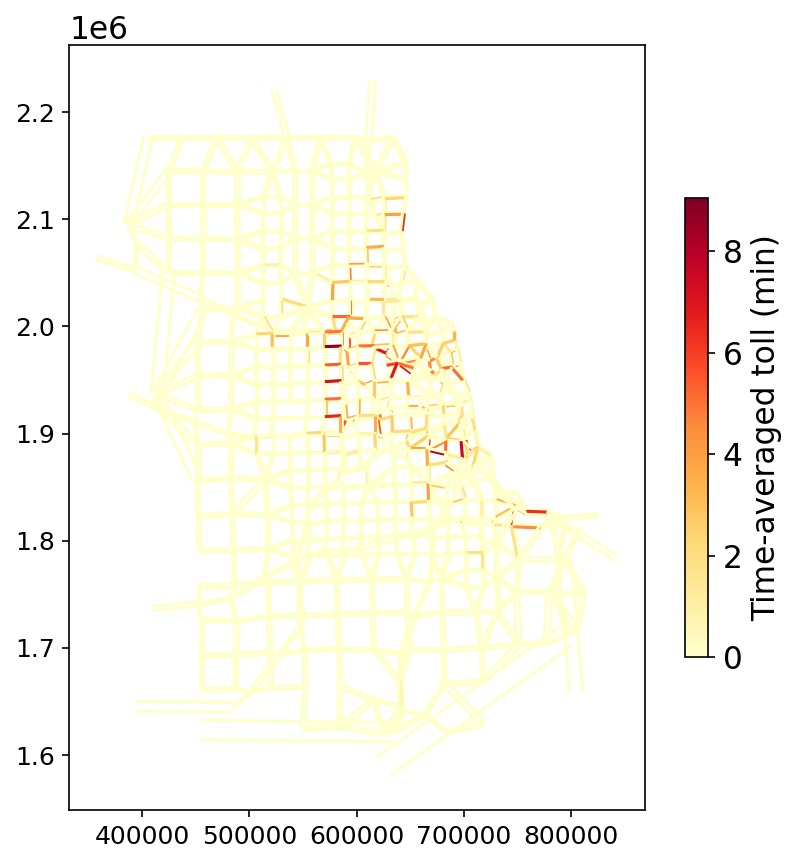}
	\caption{Average link toll in network.}
	\label{toll_link_avg}
\end{figure}

The spatial distribution of tolls is shown in \cref{toll_link_avg}.
Most of the tolls were charged inside and around the downtown congested area.
It appears that a toll pattern similar to perimeter control might have been learned by the model.

\begin{figure}[htp]
	\centering
	\subfloat[Average toll]{\includegraphics[clip, width=0.6\hsize]{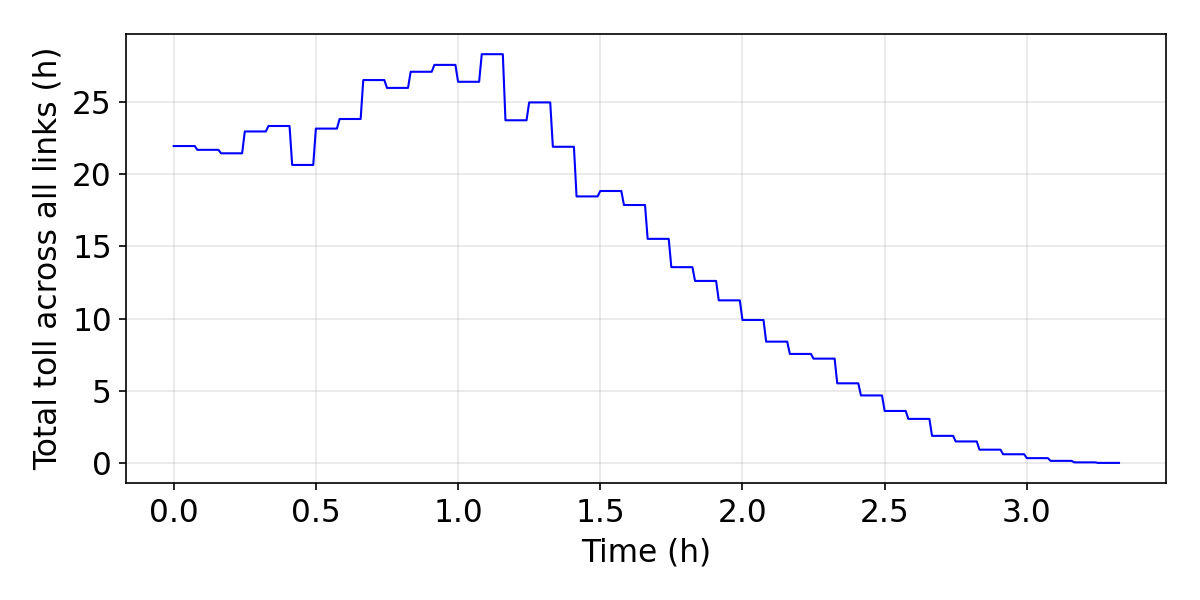}\label{toll_total_timeseries}}\\
	\subfloat[Total vehicle count]{\includegraphics[clip, width=0.6\hsize]{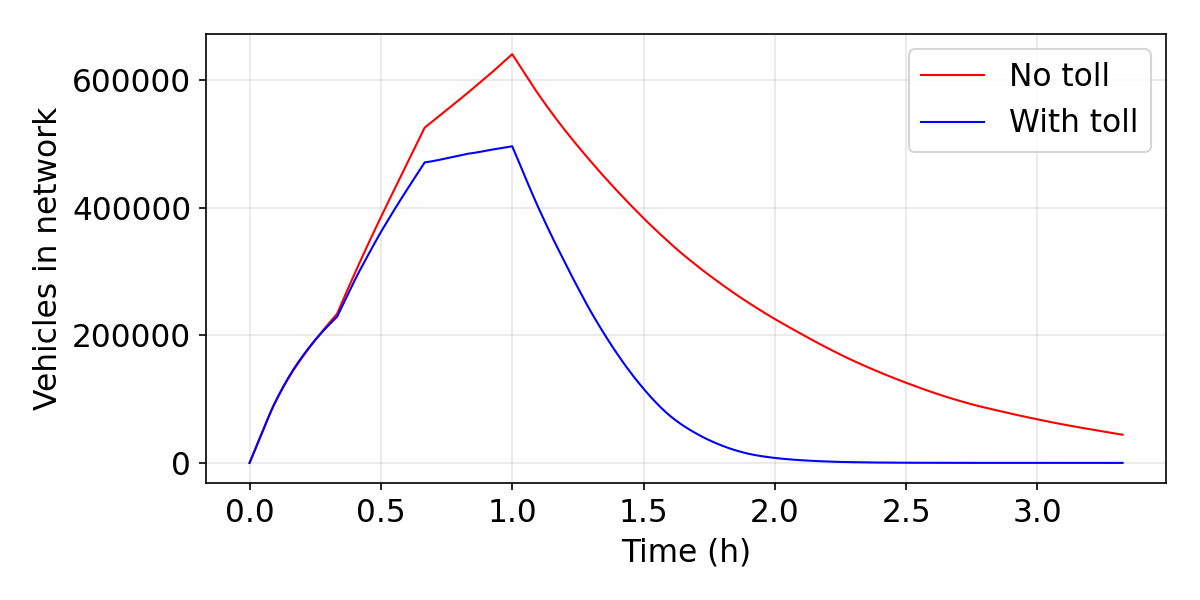}\label{toll_vehicle_count}}\\
	\subfloat[Average speed]{\includegraphics[clip, width=0.6\hsize]{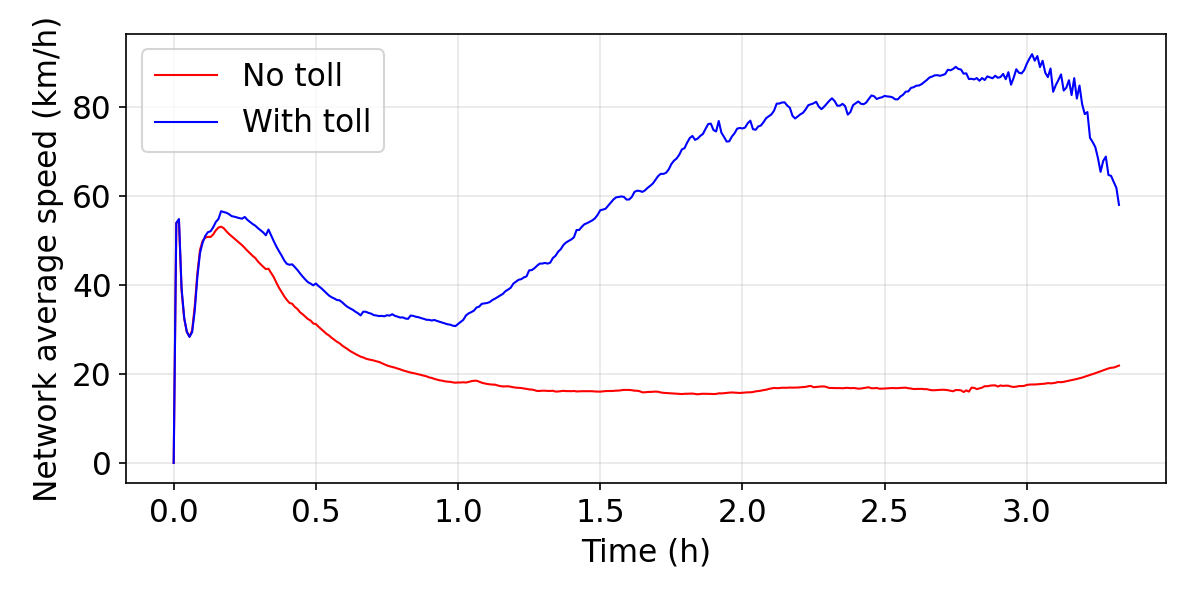}\label{toll_avg_speed}}
	\caption{Time-series of toll and traffic states.}
	\label{dynamics_network}
\end{figure}

\Cref{dynamics_network} shows the dynamics of toll and traffic state in the best pricing solution.
The toll in \cref{toll_total_timeseries} has a clear peak.
Vehicle counts in \cref{toll_vehicle_count} also show similar peak patterns.
They suggest that, as the number of vehicles increased, higher tolls were likely imposed to guide their route choices.
As a result, speed dynamics (\cref{toll_avg_speed})\footnotemark{} were very different between the with-toll and no-toll scenarios.
With the toll, congestion was avoided or mitigated, resulting in a significant improvement in speed.

\footnotetext{
	Note that \cref{toll_avg_speed} contains some anomalies (i.e., visually unusual but non-essential phenomena), such as sudden increases and decreases in speed at the initial stage ($< 0.1$ h) in both scenarios and high speed around 80 km/h after $t > 1.5$ h in the with-toll scenario.
	Although these patterns are visually noticeable, the number of vehicles during these time periods is very small, as shown in \cref{toll_vehicle_count}.
	Therefore, these patterns do not represent the overall characteristics of the network traffic.
}

\begin{figure}[htp]
	\centering
	\includegraphics[clip, width=0.5\hsize]{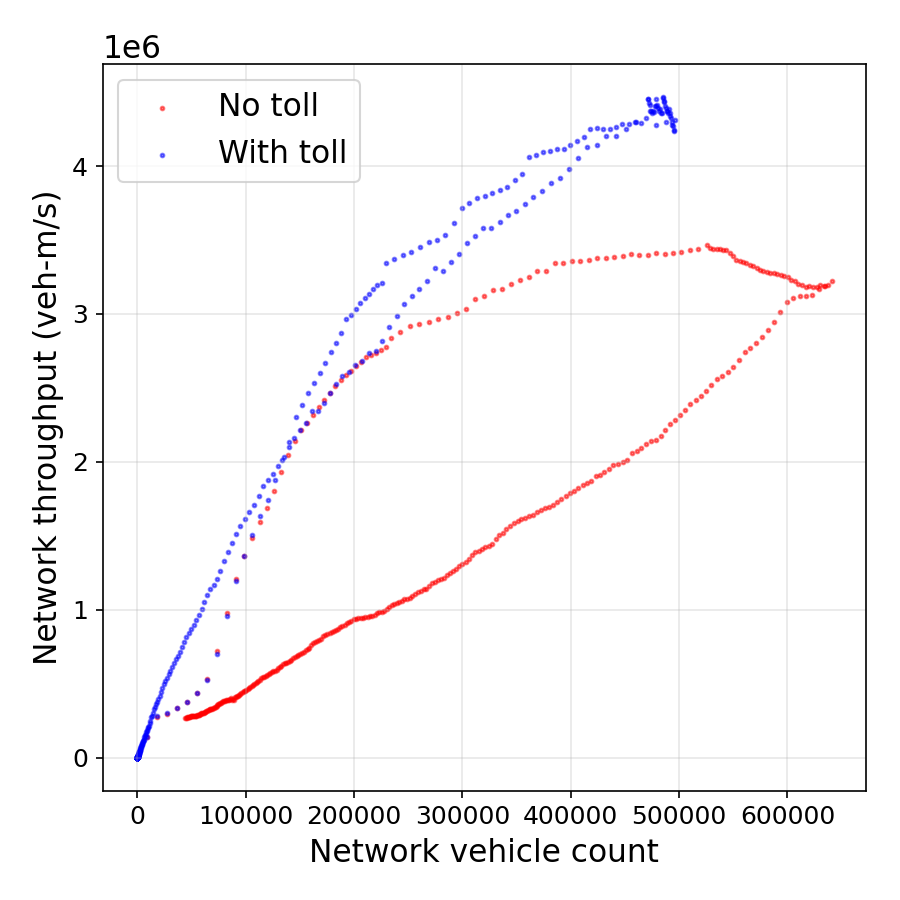}
	\caption{Macroscopic Fundamental Diagram.}
	\label{toll_mfd}
\end{figure}

Finally, in \cref{toll_mfd}, traffic dynamics are analyzed by using the Macroscopic Fundamental Diagram (MFD) \citep{mahmassani1984nfd,Geroliminis2007mfd}.
The results are highly interpretable.
In the no-toll scenario, the network state entered the congested regime of the MFD, and caused a significant hysteresis phenomenon \citep{Geroliminis2011mfdhyst,Geroliminis2011mfd}, leading to a very inefficient situation.
On the other hand, in the with-toll scenario, network traffic was routed in an efficient manner by avoiding the congested regime and significant hysteresis.
The maximum throughput was even increased.
These results suggest that the derived optimal pricing was very efficient.

\subsubsection{Comparison with SPSA}

In order to quantify the advantage of the proposed AD-based gradient computation, we solved the same congestion pricing problem using SPSA \citep{Spall1998spsa}, a conventional derivative-free optimization method in the simulation and DTA literature \citep{Balakrishna2007calib,Lu2015wspsa}.
SPSA estimates the gradient by evaluating the objective function at two symmetrically perturbed points $\tthh \pm c_k \bm{\delta}$, where each component of $\bm{\delta}$ is an independent Bernoulli $\pm 1$ random variable.
The gradient approximation at the $k$-th iteration is defined as
\begin{align}
	\hat{\bm{g}}_k = \frac{J(\tthh + c_k\bm{\delta}) - J(\tthh - c_k\bm{\delta})}{2 c_k \bm{\delta}},	\label{eq_spsa_gradient}
\end{align}
which requires only two forward simulations per iteration regardless of the parameter dimension.
Then, the parameters are updated as
\begin{align}
 	\tthh_{k+1} = \tthh_k - a_k \hat{\bm{g}}_k,
\end{align}
where the step size $a_k = a/(A+k)^{\alpha}$ and perturbation magnitude $c_k = c/k^{\gamma}$ follow the standard decay schedule recommended by \citet{Spall1998spsa} with $A=100$, $\alpha = 0.602$, $\gamma = 0.101$, and the initial parameters $c=30$ and $a=0.0001$ are calibrated to achieve the best performance as much as possible.
The objective function for SPSA is the same as that for AD, \cref{eq_toll_obj}.

In order to conduct a fair comparison, we used the same total computation time for SPSA and AD.
In AD, one iteration requires a forward simulation and reverse AD, and a total of \num{10000} iterations were performed in \num{8365} sec.
SPSA requires two forward simulations per iteration, which results in a shorter iteration time than one AD iteration.
Therefore, the total number of iterations for SPSA was set to \num{17000}, which took \num{8532} sec.

\begin{figure}[htp]
	\centering
	\subfloat[Convergence]{\includegraphics[clip, width=0.5\hsize]{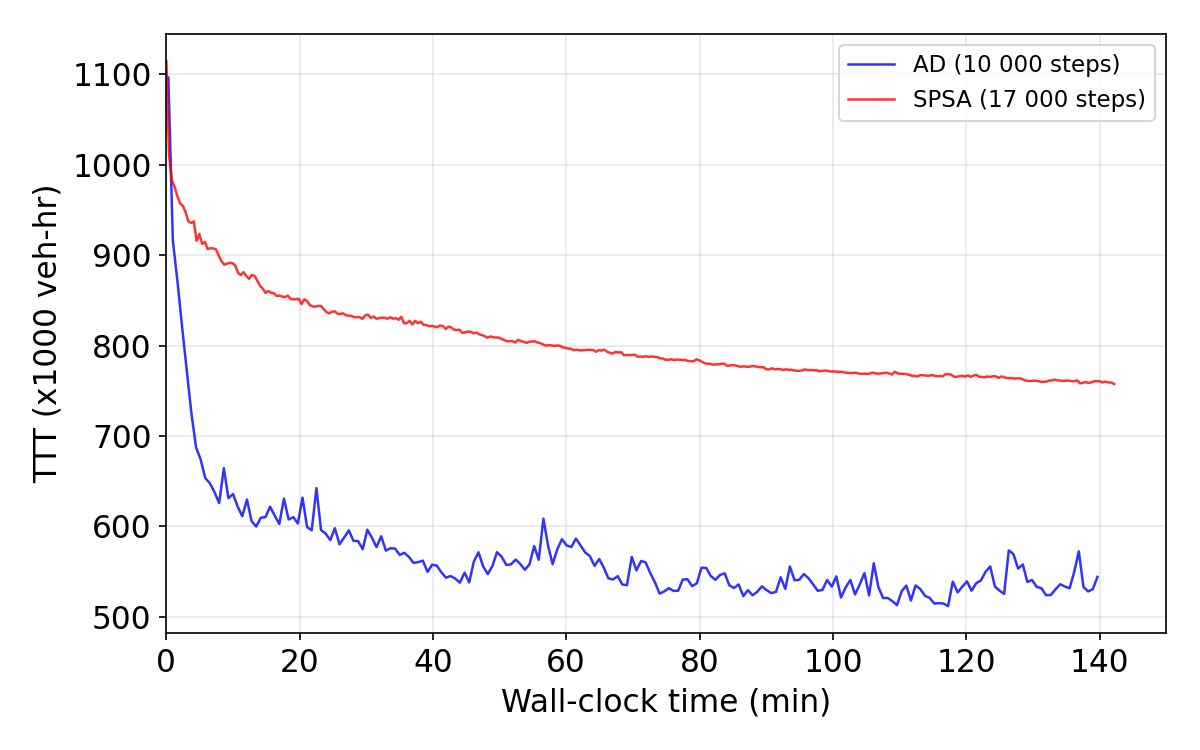}\label{ad_vs_spsa_convergence}}\\
	\subfloat[Link average toll]{\includegraphics[clip, width=0.5\hsize]{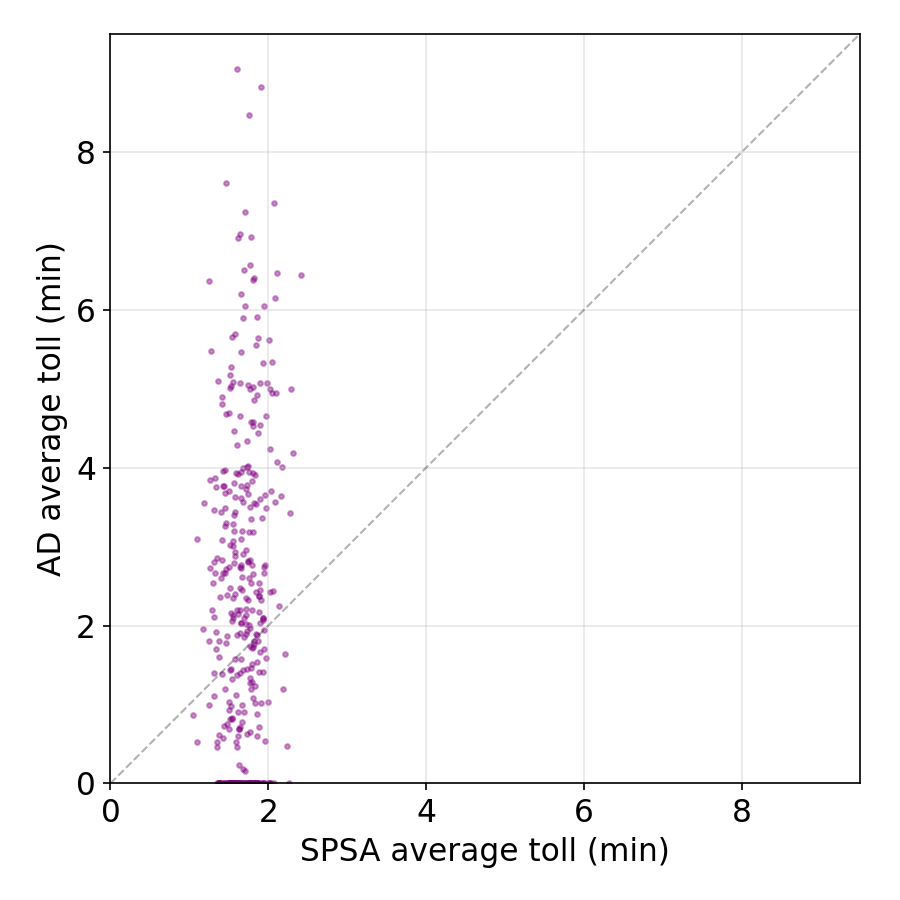}\label{ad_vs_spsa_toll_scatter}}
	\caption{Comparison between AD and SPSA.}
	\label{ad_vs_spsa}
\end{figure}

\Cref{ad_vs_spsa} summarizes the comparison results.
According to the convergence plot (\cref{ad_vs_spsa_convergence}), AD converges faster and achieves a better objective value than SPSA.
It is noteworthy that SPSA achieved non-trivial improvements, as already demonstrated in existing studies; it reduced TTT from the initial \num{1120100} veh-hr to \num{757566} veh-hr.
However, this value is still substantially larger than the AD result of \num{509743} veh-hr.

\Cref{ad_vs_spsa_toll_scatter} compares the link average tolls obtained by AD and SPSA.
A clear tendency can be observed.
While AD tailored the toll for each link, SPSA assigned similar tolls to all links.
This demonstrates the differential capability of AD.
Namely, while SPSA evaluated only the averaged effects over all links captured by the aggregated gradient approximation in \cref{eq_spsa_gradient}, AD accurately distinguished the impact of each link toll on network traffic by using per-parameter exact partial derivatives.

\section{Conclusion}\label{sec_conclusion}

In this study, an end-to-end differentiable framework based on automatic differentiation for dynamic network simulation is presented.
The model is based on the Link Transmission Model and the Dynamic User Optimum route choice model, which can be considered standard and reasonable models in transportation research.
By leveraging the nature of these models, the proposed framework is formulated as differentiable almost everywhere without significant approximations; this is a substantial advantage compared with existing differentiable traffic simulation models.
Extensive numerical verifications in small interpretable toy networks confirmed that the model computes qualitatively and quantitatively reasonable partial derivatives.
Furthermore, the model was applied to a dynamic congestion pricing optimization problem on the Chicago-Sketch network to demonstrate its overall accuracy and scalability.

The proposed simulator, {\it UNsim}, is implemented using Python and JAX and released as open-source software on GitHub (\url{https://github.com/toruseo/UNsim}) and PyPI ({\tt pip install unsim}).
All code that reproduces the presented results is also published in the same GitHub repository.

For future research, the following directions are worth considering.
First, there is room for improvement in the model and algorithms to facilitate numerical applications.
During this study's numerical experiments, it was found that the proposed model frequently produces zero gradients.
This is a theoretically reasonable behavior: the information flow in the LWR model with a triangular FD is quite asymmetric (e.g., information in free-flowing traffic only propagates downstream; traffic capacity in free-flowing traffic does not have any impact on traffic), and the proposed differentiable model accurately and exactly reproduces these behaviors.
However, such zero gradients are sometimes troublesome for gradient-based optimization algorithms, as they may create a large plateau on the objective function.
The development of tailored optimization algorithms or the introduction of artificial smoothing or viscosity methods might be worth considering for numerical convenience.
In addition, computation cost and memory consumption can be further reduced by employing existing techniques in automatic differentiation.

Second, the application of the proposed framework to other tasks is important.
Since the primary contribution of this study is the development of the differentiable simulation model, the numerical examples focused on dynamic congestion pricing as a representative application.
Nevertheless, the same gradient-based approach is directly applicable to a broader set of transportation engineering tasks, such as OD demand estimation (e.g., by minimizing the difference between observed and simulated flow with an assumed demand), FD parameter calibration (e.g., similar to OD estimation), network design optimization (e.g., by adding link capacity decision variables and budget constraints), and other traffic control and optimization (e.g., by adding traffic signal configuration or ramp-metering capacity as decision variables and adjusting the node model to account for them).
The proposed model can directly incorporate these problems.
Furthermore, since the computation time of the simulator is short, application to real-time control would also be promising.

Third, direct integration with deep learning would be interesting.
Since the simulator is end-to-end differentiable, neural network components can be embedded directly into the computation graph and trained jointly with the physical model parameters via the same gradient pipeline, as demonstrated in other transportation studies \citep{Sifringer2020behavior,Shi2022tse,Liu2023network,Ma2025route_arxiv}.
In order to exploit this capability, physics-informed deep learning integration is a promising direction: for example, parameterizing spatiotemporal OD demand patterns or fundamental diagram functions with neural networks while retaining the LTM and DUO as the physical backbone.
Such hybrid models would combine data-driven flexibility with the traffic-theoretical consistency guaranteed by the network traffic flow theory.

\section*{Acknowledgements}\label{sec_ack}

Part of this work was financially supported by JSPS KAKENHI Grants-in-Aid 24K01002 and 25H00751 and JICA/JST SATREPS JP-
MJSA2405.
The author would also like to thank Mr.~Zhengyou Han, Prof.~Masaki Ito, and \citet{Makinoshima2026trafficsim} for providing significant inspiration for differentiable traffic flow simulations, as well as the participants of the ``Workshop: AI and Transportation Research'' held at the University of Tokyo (especially the organizer, Prof.~Takamasa Iryo) for motivating this approach.

\section*{Declaration of competing interests}

There are no conflict of interests.

\section*{Declaration of generative AI use}

During the preparation of this work the author used Claude and GPT AI families for coding and writing assistance.
After using this tool/service, the author reviewed and edited the content as needed and take full responsibility for the content of the published article.

\section*{CRediT Author Contributions}

TS did everything.


\appendix
\renewcommand{\theequation}{\thesection.\arabic{equation}}
\renewcommand{\thefigure}{\thesection.\arabic{figure}}
\renewcommand{\thetable}{\thesection.\arabic{table}}
\setcounter{equation}{0}
\setcounter{figure}{0}
\setcounter{table}{0}

\section{Automatic differentiation and computation graphs}\label{apx_ad}

This appendix provides a brief overview of automatic differentiation (AD) and its important concept, the {\it computation graph}, and explains why AD is beneficial for traffic simulation.
For a comprehensive treatment, see \citet{Griewank2008ad} and \citet{Baydin2018adsurvey}.

A function $J(\tthh)$ given by a fixed finite evaluation procedure can be written as a composition of elementary operations (e.g., addition, multiplication, $\min$, $\exp$, $\log$).
This decomposition can be represented as a {\it computation graph}: a directed acyclic graph in which each node corresponds to an intermediate variable and each edge represents a dependency.
AD computes derivatives of $J$ by systematically applying the chain rule along this graph.

\begin{figure}[htbp]
	\centering
%
%
%
%
%
%
	\includegraphics[clip, width=0.7\hsize]{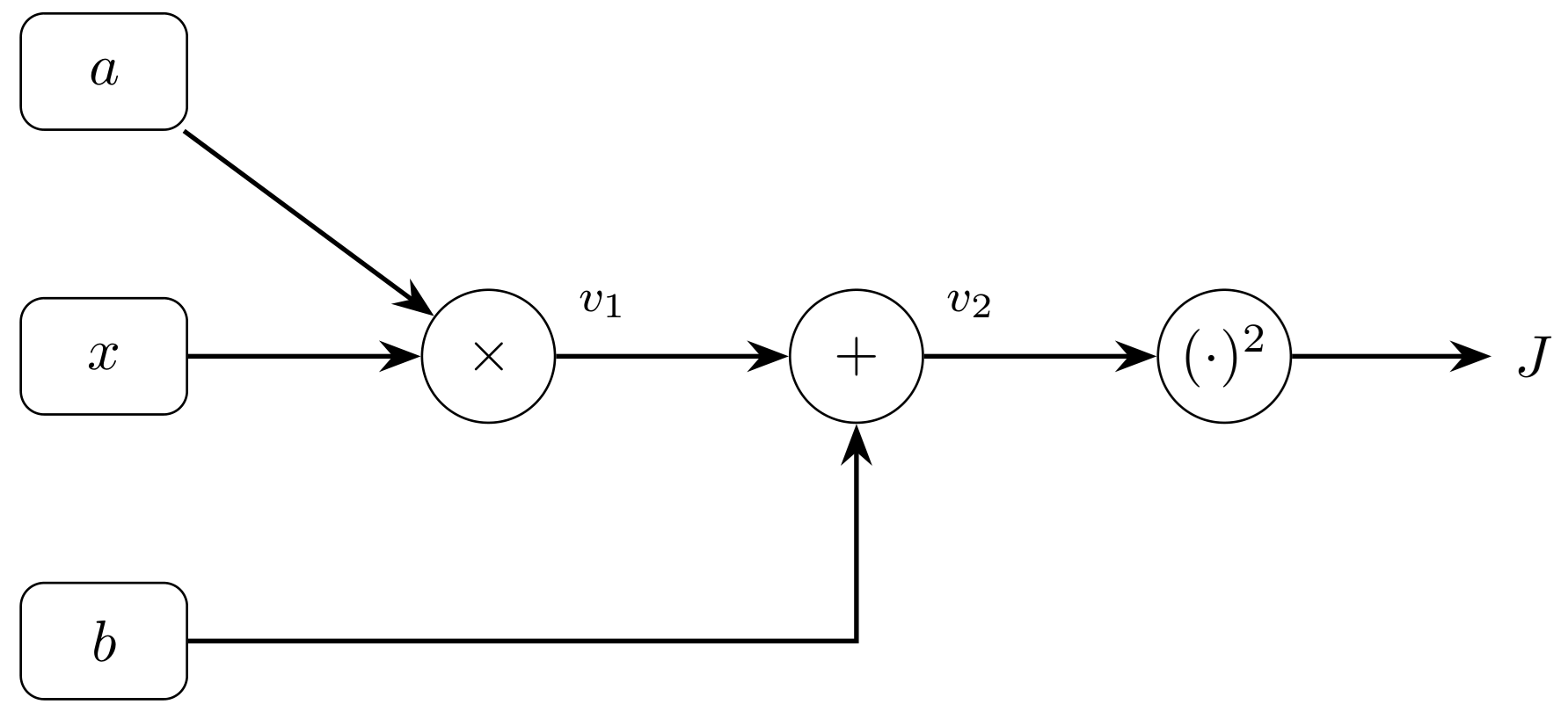}
	
	\caption{Computation graph for $J = (ax+b)^2$.}
	\label{fig_compgraph}
\end{figure}

As a minimal example, consider the function $J = (a x + b)^2$ with parameters $\tthh = (a, b)$ and a fixed input $x$.
The computation can be decomposed into three elementary operations:
\begin{align}
	v_1 = a x, \quad v_2 = v_1 + b, \quad J = v_2^2.	\label{eq_ad_example}
\end{align}
This can be represented as a computation graph shown in \cref{fig_compgraph}.

In {\it forward mode}, one propagates tangent values from inputs to output.
For instance, setting $\dot{a} = 1$ and $\dot{b} = 0$ yields $\dot{v}_1 = x$, $\dot{v}_2 = x$, and $\dot{J} = 2 v_2 x = 2(ax+b)x$, which is $\del J / \del a$.
A second pass with $\dot{a} = 0$, $\dot{b} = 1$ gives $\del J / \del b = 2(ax+b)$.
In {\it reverse mode}, one propagates adjoints from output to inputs in a single pass: starting with $\bar{J} = 1$, one obtains $\bar{v}_2 = 2 v_2$, then $\bar{v}_1 = \bar{v}_2 = 2v_2$ and $\bar{b} = \bar{v}_2 = 2v_2$, and finally $\bar{a} = \bar{v}_1 \cdot x = 2v_2 x$.
Both derivatives are obtained simultaneously, regardless of the number of parameters.

More generally, in forward mode, the graph is traversed from inputs to outputs, propagating a tangent vector $\dot{\tthh}$ to obtain the directional derivative $\dot{J} = (\del J / \del \tthh) \dot{\tthh}$.
The cost is proportional to one forward evaluation per input direction, so computing the full gradient with respect to $P$ parameters requires $P$ passes.
In reverse mode, the graph is traversed from outputs to inputs, propagating an adjoint backward to obtain the full gradient $\del J / \del \tthh \in \R^P$ in a single pass, regardless of $P$.
This is the same mechanism as backpropagation in neural networks.

The distinction between forward and reverse mode has a direct practical consequence for traffic simulation.
A conventional (non-differentiable) simulator treats $J$ as a black box; sensitivity analysis requires the finite-difference method, which perturbs each parameter individually, such as the central difference
\begin{align}
	\widehat{\frac{\del J}{\del x}} = \frac{J(x+\epsilon)-J(x-\epsilon)}{2\epsilon}		\label{eq_central}
\end{align}
for a particular parameter $x$.
In order to obtain all of the partial derivatives, this costs $O(P)$ simulation runs.
In contrast, a differentiable simulator exposes the computation graph to an AD framework.
Reverse-mode AD then computes the exact gradient at a cost that is a small constant multiple ($2$--$5\times$) of one forward simulation, regardless of $P$.
This makes gradient-based optimization feasible even for high-dimensional problems such as time-dependent and link-dependent congestion pricing across a large network.
The gradient itself is also a useful output for sensitivity analysis, as it reveals which parameters have the greatest influence on the objective.

For a traffic simulation with $T_S$ timesteps, the computation graph is the sequential composition $\xx_0 \to \xx_1 \to \cdots \to \xx_{T_S}$, where each transition $\xx_{t+1} = g(\xx_t, t; \tthh)$ contributes a subgraph.
Reverse-mode AD traverses this chain backward, accumulating $\del J / \del \tthh$ in $O(T_S)$ time.
Two requirements must be met: (i) every operation in $g$ must be (sub)differentiable, and (ii) the graph structure must be static and independent of the parameter values, so that the AD framework can trace it once and reuse the trace.
How the proposed simulator satisfies both requirements is described in \cref{sec_diff}.

The example in \cref{fig_compgraph} involves only three operations, but the principle scales without modification.
A traffic simulation comprising thousands of timesteps, each with demand/supply computations, node models, route choice, and cumulative count updates, forms a computation graph with millions of nodes when properly modeled and implemented.
As long as every elementary operation in the graph is (sub)differentiable, the AD framework applies the chain rule automatically across the entire graph, and the user need not derive any gradient formula by hand or apply numerical differentiation.
This is the fundamental advantage of the differentiable simulation approach adopted in this study.

The challenge is that $J$ must be expressed as a {\it fixed finite evaluation}, as noted earlier.
In particular, operations such as for-loops and if-then rules, which frequently appear in standard traffic simulation methods, are not compatible with AD, because they change the number of evaluations endogenously.
One of the main contributions of this work is to reformulate the simulation model into a form that admits fixed finite evaluation, without sacrificing its ability to represent traffic flow dynamics.

%


\end{document}